\documentclass{IEEEoj}

\usepackage[active]{srcltx} 
\usepackage[english]{babel}
\usepackage{booktabs}
\usepackage[cmex10]{amsmath}
\usepackage{amsfonts}
\usepackage[final]{graphics}
\usepackage[final]{graphicx}
\usepackage{amsbsy}
\usepackage{amsbsy}
\usepackage{amssymb}
\usepackage{url}
\usepackage{enumerate}
\usepackage{cite}
\usepackage{psfrag}
\usepackage{color}
\usepackage{euscript}
\usepackage[utf8]{inputenc}
\usepackage{algorithmic}
\usepackage[font=footnotesize]{caption}
\usepackage{subcaption} 
\usepackage[ruled,vlined]{algorithm2e}
\usepackage{stfloats}
\usepackage{pifont}

\newcommand{\Cset}{\mathbb{C}}

\renewcommand{\OJlogo}{}

\newcommand{\bm}[1]{{\mathbf{#1}}}
\newcommand{\w}{\bm{w}}

\renewcommand{\H}{\bm{H}}
\newcommand{\h}{\bm{h}}
\renewcommand{\a}{\bm{a}}   

\newcommand{\I}{\bm{I}}

\newcommand{\db}{\bm d}

\newcommand{\Cb}{\bm C}

\newcommand{\Rb}{\bm R}

\newcommand{\nb}{\bm n}

\newcommand{\eqdef}{\triangleq}

\newcommand{\herm}{\text{H}}
\newcommand{\trasp}{\text{T}}

\newcommand{\capa}{\EuScript{R}}

\def\bdm#1\edm{\begin{displaymath}#1\end{displaymath}}
\def\be#1\ee{\begin{equation}#1\end{equation}}
\def\barr#1\earr{\begin{align}#1\end{align}}

\begin{document}

\def\receiveddate#1{}
\def\reviseddate#1{}
\def\accepteddate#1{}
\def\publisheddate#1{}
\def\doiinfo#1{}


\title{A Robust Two-Stage Protocol for STAR-RIS-Aided ISAC Networks: Joint Beamforming and Mode Optimization}

\author{Ziming~Liu\authorrefmark{1} (MEMBER, IEEE),
Tao~Chen\authorrefmark{1} (MEMBER, IEEE),
Giacinto~Gelli\authorrefmark{2} (SENIOR MEMBER, IEEE), 
Vincenzo Galdi\authorrefmark{3} (FELLOW, IEEE), \\
AND Francesco~Verde\authorrefmark{4} (SENIOR MEMBER, IEEE)
}
\affil{College of Information and Communication Engineering of Harbin Engineering University, 
Harbin, 150001 China}
\affil{Department of Electrical Engineering and
Information Technology, University Federico II, I-80125 Naples, Italy}
\affil{Fields \& Waves Lab, Department of Engineering, University of Sannio,  
I-82100 Benevento, Italy}
\affil{Department of Engineering, University of Campania Luigi Vanvitelli, I-81031 Aversa, Italy}
\corresp{CORRESPONDING AUTHOR: F.~VERDE (e-mail: francesco.verde@unicampania.it).}
\authornote{}
\markboth{A Robust Two-Stage Protocol for STAR-RIS-Aided ISAC Networks: Joint Beamforming and Mode Optimization}{Liu \textit{et al.}}

\begin{abstract}

This paper investigates the robust design of integrated sensing 
and communication (ISAC) systems assisted by simultaneously 
transmitting and reflecting reconfigurable intelligent surfaces 
(STAR-RISs), which act as multi-functional programmable 
metasurfaces enabling concurrent sensing and communication 
over the entire coverage area (full space operation). 
To exploit the dual transmission-reflection capability of 
STAR-RISs in practical deployments, we propose a two-stage 
ISAC protocol operating on a slot-by-slot basis: a preparation 
phase jointly performs direction-of-arrival (DoA) estimation 
for outdoor users located in the reflection space and downlink 
communication to both outdoor and indoor users, while a 
subsequent communication phase leverages the acquired angular 
information to further enhance downlink communication.
To capture sensing uncertainty as well as imperfect channel knowledge, 
the DoAs of outdoor users are modeled as Gaussian random 
variables, and the non-line-of-sight (NLoS) channel components 
of outdoor links are characterized through their spatial 
covariance statistics rather than assumed perfectly known, 
enabling a robust design that incorporates average 
communication performance into the optimization.
Building on this framework, we formulate a performance-balanced 
optimization problem that maximizes the communication sum-rate
(i.e., the throughput) while guaranteeing the required sensing accuracy, jointly 
determining the beamforming vectors at the base station, 
the STAR-RIS transmission and reflection coefficients in both 
stages, and the metasurface partition between energy-splitting 
and transmit-only modes, while explicitly enforcing the 
physical constraints of STAR-RIS elements and sensing-quality 
requirements.
The resulting non-convex mixed discrete-continuous problem 
poses significant algorithmic challenges, due to the coupling 
among beamforming variables, STAR-RIS physical constraints, 
and the binary mode-selection vector. To address this, we 
develop a tailored alternating optimization framework that 
jointly handles the continuous and combinatorial components 
of the problem within a unified iterative procedure, with 
proven monotonic convergence to a stationary point.
Numerical results demonstrate that the proposed robust design 
achieves an effective sensing-communication trade-off, yielding approximately $15\%$ 
throughput gain over the most competitive benchmark scheme that neglects NLoS statistical 
channel characterization, with robustness maintained even under DoA estimation errors 
and imperfect NLoS channel knowledge.

\end{abstract}


\maketitle

\section{Introduction}

\IEEEPARstart{M}{etasurfaces} are artificial electromagnetic (EM) 
structures composed of subwavelength elements (``meta-atoms'') 
integrated with tunable microelectronic components, 
such as diodes and varactors, which enable programmable control 
over the amplitude, phase, and polarization of incident waves 
\cite{Holloway.2012}. When empowered with reconfigurability, 
these structures give rise to {\em reconfigurable intelligent 
surfaces (RISs)}, which emerged as a key enabler for 
next-generation wireless networks due to their ability to 
reshape the propagation environment in a software-defined 
manner \cite{Renzo.2020,Kudathanthirige.2020.ICC,Tang.2021.TWC}. 
By dynamically manipulating the reflected or transmitted 
wavefronts, RISs allow the establishment of favorable 
communication links, the mitigation of blockages, and the enhancement 
of coverage in complex propagation scenarios 
\cite{Cui.2014,Huang.2017,Liu.2023}. Owing to their low power 
consumption, compact form factor, and deployment flexibility, 
RISs are particularly attractive for integration into urban 
infrastructures, vehicles, and buildings, thus supporting 
the vision of environment-aware and user-centric wireless 
networks \cite{Meng.2024.TVT,Li.2024.TOC,Wang.2025.JSAC,
Yigit.2025.TOC,Zappia.2026}.

\IEEEpubidadjcol

Despite their advantages, conventional RIS architectures 
typically rely on \textit{reflective} elements, which inherently limit 
their angular coverage to a single half-space. This constraint 
restricts their applicability in scenarios where users are 
distributed on both sides of the surface. To overcome this 
limitation, {\em simultaneously transmitting and reflecting 
reconfigurable intelligent surfaces (STAR-RISs)} have been 
recently proposed \cite{Bao.2021.TWC,Mu.2022.TWC,Xu.2021.TWC}. 
By leveraging advanced meta-atom designs and interlayer 
structures, STAR-RISs are capable of concurrently reflecting 
and transmitting incident signals, thereby enabling \textit{full-space} 
EM control \cite{Xu.2021CM,Hu.2022}. This unique capability 
makes STAR-RISs a promising solution for supporting 
heterogeneous users located in distinct spatial regions, 
such as indoor and outdoor environments separated by a 
building facade.

STAR-RISs can operate under different \textit{working modes}, 
including energy splitting (ES), mode switching (MS), time 
division (TD), polarization division (PD), and frequency 
division (FD) \cite{Verde.2024.SPM,Liu.2021.}. Among these, 
the ES mode has received the most attention in the literature, 
due to its ability to simultaneously support transmission and 
reflection while respecting energy conservation principles
\cite{Zhong.2022,Xu.2022,Zhu.2024,Liu.2022.ICC,
Huang.2025.IOTJ,Wang.2023.WCL,Wu.2024.CL}. In this mode, 
the transmitted and reflected signals are inherently 
coupled through both amplitude and phase constraints. 
Other modes, such as MS and TD \cite{Ju.2024.TOC}, can be 
interpreted as special cases or extensions of the ES mode 
with additional constraints, while PD and FD modes impose stringent 
requirements on signal polarization or frequency selectivity 
\cite{Cai.2017,Wang.2024.Sci}, and have therefore received 
limited attention in the literature.

\subsection{Motivations}

In parallel with the development of programmable 
metasurfaces, {\em integrated sensing and 
communication (ISAC)} has emerged as a fundamental 
paradigm for sixth-generation (6G) wireless networks 
\cite{Cui.2021,Dong.2025}. ISAC aims to unify sensing 
and communication functionalities within a single 
system, enabling spectrum- and hardware-efficient 
operation. This dual functionality is particularly 
relevant for smart city deployments, where base 
stations (BSs) must simultaneously serve mobile users and 
monitor the surrounding environment 
\cite{Mehmood.2017.CommMag,Darsena.2023}. STAR-RISs 
are especially well suited for ISAC applications, as 
their full-space control capability allows simultaneous 
information delivery and environmental sensing in 
physically separated spatial regions --- a distinctive 
advantage over conventional reflective-only RIS 
architectures, which are inherently confined to a 
single half-space.

Despite this promise, two fundamental challenges limit 
the practical deployment of STAR-RIS-aided ISAC 
systems. First, in dynamic outdoor environments, the 
DoAs of mobile users are not perfectly known 
and must be inferred through sensing mechanisms 
operating on a slot-by-slot basis. This sensing 
uncertainty propagates into the communication design, 
degrading beamforming accuracy if not explicitly 
accounted for. Second, the line-of-sight (LoS) and non-line-of-sight 
(NLoS) channel components of outdoor links
are difficult to 
estimate instantaneously due to the complexity of the 
scattering environment and the associated pilot 
overhead. Assuming perfect NLoS knowledge --- as done 
in most existing works --- leads to overly optimistic 
performance predictions and suboptimal designs in 
realistic deployments. These two challenges motivate 
the development of a robust STAR-RIS-aided ISAC 
framework that explicitly accounts for both 
direction-of-arrival (DoA) estimation uncertainty 
and imperfect or partial NLoS channel knowledge.

\subsection{Literature review}
Herein, we provide a structured review of the 
literature most closely related to the present work, 
organized along three main research threads: 
STAR-RIS-aided ISAC systems, DoA estimation in 
RIS/STAR-RIS-assisted networks, and statistical channel state information (CSI) 
design for robust beamforming. For each thread, 
the key contributions of existing works are discussed 
and their limitations with respect to 
the scenario addressed in this paper are highlighted.

\subsubsection{STAR-RIS-aided ISAC systems}

The integration of STAR-RIS into ISAC systems has 
attracted growing research interest. The 
sensing-at-STAR-RIS architecture, where dedicated 
sensors mounted on the metasurface perform target 
detection while the BS serves indoor 
and outdoor users simultaneously, was pioneered 
in~\cite{Wang.2023.TWC}; however, that design assumes 
perfect knowledge of all channel components and does 
not consider estimation uncertainty. Active STAR-RIS 
configurations operating in full-duplex mode were 
considered in~\cite{Zhang.2024.TWC}, where transmit 
and receive beamforming are jointly optimized under 
radar signal-to-interference-plus-noise ratio (SINR) 
constraints, again under idealized CSI assumptions. 
Multi-target multi-user 
STAR-RIS-ISAC systems were addressed 
in~\cite{Zhang.2025} via signature sequence modulation 
to disambiguate targets sharing the same angular 
DoA relative to the BS, with focus on beampattern 
gain optimization rather than robust design 
under uncertainty. A STAR-RIS with movable elements 
for covert ISAC was studied in~\cite{Zhou.2025}, 
where element deployment and beamforming are jointly 
optimized within a single transmission stage, without 
protocol-level design. A hybrid STAR-RIS architecture 
combining active transmissive and passive reflective 
elements was proposed in~\cite{Yigit.2025.TOC} to 
enable full-space multi-target sensing and 
communication, without accounting for DoA estimation 
uncertainty or NLoS channel imperfections. 
High-mobility millimeter-wave scenarios were addressed 
in~\cite{Li.2024.TOC}, where an active STAR-RIS 
supports dynamic scatterer tracking; however, that 
approach does not consider two-stage protocol design 
or statistical NLoS channel characterization.

From a different perspective, a robust RIS-aided 
scheme for securing a full-duplex ISAC network was 
proposed in~\cite{Illi.2026.TWC}, where phase shifts, 
transmit beamforming, artificial noise covariance, 
and uplink combining vectors are jointly optimized to 
maximize sensing performance under secrecy 
constraints; however, that work relies on conventional 
reflective-only RIS and does not exploit the 
indoor/outdoor coverage duality enabled by STAR-RIS. 
Path interference --- arising from the combination of 
direct-path and reflected-path interference --- was 
identified as a critical impairment in bistatic 
RIS-ISAC systems with passive radar receivers 
in~\cite{Bazzi.2025.JSAC}, where an optimization 
framework is proposed to minimize its impact while 
satisfying communication and radar 
signal-to-noise-plus-distortion ratio (SNDR) 
requirements; that scenario, however, differs 
significantly from the present work, as it considers 
a bistatic configuration without STAR-RIS 
and does not address multi-user beamforming design 
or DoA estimation uncertainty.

\subsubsection{Direction-of-Arrival estimation 
in RIS/STAR-RIS systems}

DoA estimation accuracy in metasurface-assisted 
systems is fundamentally limited by noise, propagation 
conditions, and hardware constraints. STAR-RIS-aided 
simultaneous indoor and outdoor three-dimensional 
localization was studied in~\cite{He.2023.IET}, where 
Fisher information analyses and Cram\'{e}r-Rao lower 
bounds (CRLBs) characterize the fundamental 
performance limits; that work focuses on localization 
accuracy bounds without addressing joint beamforming 
optimization or robust design under NLoS uncertainty. 
Full-space DoA estimation assisted by STAR-RIS was 
investigated in~\cite{Li.2024.ICMMT}, demonstrating 
the feasibility of simultaneous indoor and outdoor 
DoA estimation. Robust RIS-based DoA estimation 
algorithms under mixed amplitude and phase constraints 
were developed in~\cite{Li.2024.SPL}. A single-antenna 
DoA estimation scheme enabled by RIS phase diversity 
was proposed in~\cite{Tian.2025.TOC}, reducing 
hardware complexity at the sensing receiver. 
Location-aware beam training and channel estimation 
for RIS-aided millimeter-wave systems via atomic norm 
minimization were addressed in~\cite{Chung.2024.TWC}. 
Collectively, these works demonstrate that DoA 
estimation errors constitute an unavoidable source 
of uncertainty in practical STAR-RIS-aided ISAC 
systems, motivating their explicit incorporation 
into the system design --- an aspect that has received 
limited attention in the existing STAR-RIS-ISAC 
literature.

\subsubsection{Statistical channel state information 
and robust beamforming}

The use of statistical CSI for robust beamforming 
in RIS-assisted systems has been studied in several 
recent works. STAR-RIS-aided ISAC beamforming design 
under statistical CSI was investigated 
in~\cite{Liu.2025.TVT}, where BS beamforming and 
STAR-RIS coefficients are optimized based on long-term 
channel statistics; however, that framework assumes 
a single transmission stage and does not account for 
the feedback of estimated DoAs across protocol 
phases. The achievable rate of STAR-RIS-assisted 
massive multiple-input multiple-output (MIMO) systems 
under spatially correlated channels was analyzed 
in~\cite{Papazafeiropoulos.2024.TWC}, characterizing 
the impact of spatial covariance on system 
performance; that analysis is purely 
communication-oriented and does not address sensing. 
In contrast to these works, the present paper develops 
a statistical design framework that jointly accounts 
for DoA estimation uncertainty through Gaussian error 
modeling and partial NLoS knowledge through spatial 
covariance characterization, within a two-stage 
protocol that explicitly coordinates the sensing and 
communication phases.

\subsection{Contributions}

Against the aforementioned background, this paper makes the 
following contributions:

\begin{itemize}

\item \textbf{Two-stage ISAC protocol design:} 
We propose a two-stage ISAC protocol for STAR-RIS-aided 
networks operating on a slot-by-slot basis. The 
preparation phase jointly performs DoA estimation of 
outdoor users and downlink communication to all users, 
while the subsequent communication phase exploits the 
acquired angular information 
to further refine 
downlink transmission. 
Unlike existing single-stage designs 
\cite{Wang.2023.TWC,Zhang.2024.TWC,Zhang.2025,
Zhou.2025}, the proposed protocol explicitly accounts 
for the distinct roles of sensing- and 
communication-oriented signaling across multiple 
phases within a slot.

\item \textbf{Robust statistical design under partial 
channel knowledge:} To capture sensing uncertainty, 
the DoAs of outdoor users are modeled as Gaussian 
random variables whose variances are determined by the 
estimation accuracy. Crucially, unlike most existing 
works that assume perfect instantaneous NLoS channel 
knowledge \cite{Wang.2023.TWC,Zhang.2024.TWC}, the 
NLoS components of outdoor links are characterized 
through their long-term spatial covariance statistics, 
enabling a robust design that incorporates average 
communication performance limits into optimization 
without requiring instantaneous NLoS CSI.

\item \textbf{Joint beamforming and mode optimization:} 
We formulate a performance-balanced optimization 
problem, which jointly designs the BS beamforming 
vectors, the STAR-RIS transmission and reflection 
coefficients in both stages, and the metasurface 
partition between energy-splitting (ES) and 
transmit-only (TO) modes. The physical feasibility 
constraints of STAR-RIS elements -- including energy 
conservation and phase coupling -- are explicitly 
enforced, together with sensing-quality requirements.

\item \textbf{Efficient algorithmic solution with 
convergence guarantees:} To tackle the resulting 
non-convex mixed discrete--continuous optimization 
problem, we develop a tailored alternating optimization 
framework, which jointly handles the continuous and 
combinatorial components within a unified iterative 
procedure, with proven monotonic convergence to a 
stationary point. The computational complexity of the 
proposed algorithm is characterized in closed form.

\item \textbf{Performance validation:} Numerical 
results demonstrate that the proposed robust design 
achieves an effective sensing-communication 
trade-off, yielding approximately $15\%$ throughput 
gain over the most competitive benchmark -- a scheme 
that shares the same optimization framework but neglects 
NLoS spatial covariance statistics -- with robustness maintained 
under significant DoA estimation errors and imperfect NLoS channel knowledge, 
and sensing constraints satisfied under all considered sensing-parameter fluctuations.

\end{itemize}

\subsection{Organization}

The remainder of the paper is organized as follows. 
Section~\ref{sec:system} presents the system model 
and the basic assumptions considered throughout the 
paper, including the STAR-RIS architecture, the 
two-stage ISAC protocol, and the DoA uncertainty 
model for outdoor users. 
Section~\ref{sec:optmization} formulates the 
optimization problem and derives the proposed 
algorithmic solution. 
Section~\ref{sec:Simulation} provides numerical 
results and performance comparisons. 
Section~\ref{sec:conclusions} concludes the paper.

\begin{figure}[t]
    \centering
    \includegraphics[width=1\linewidth]{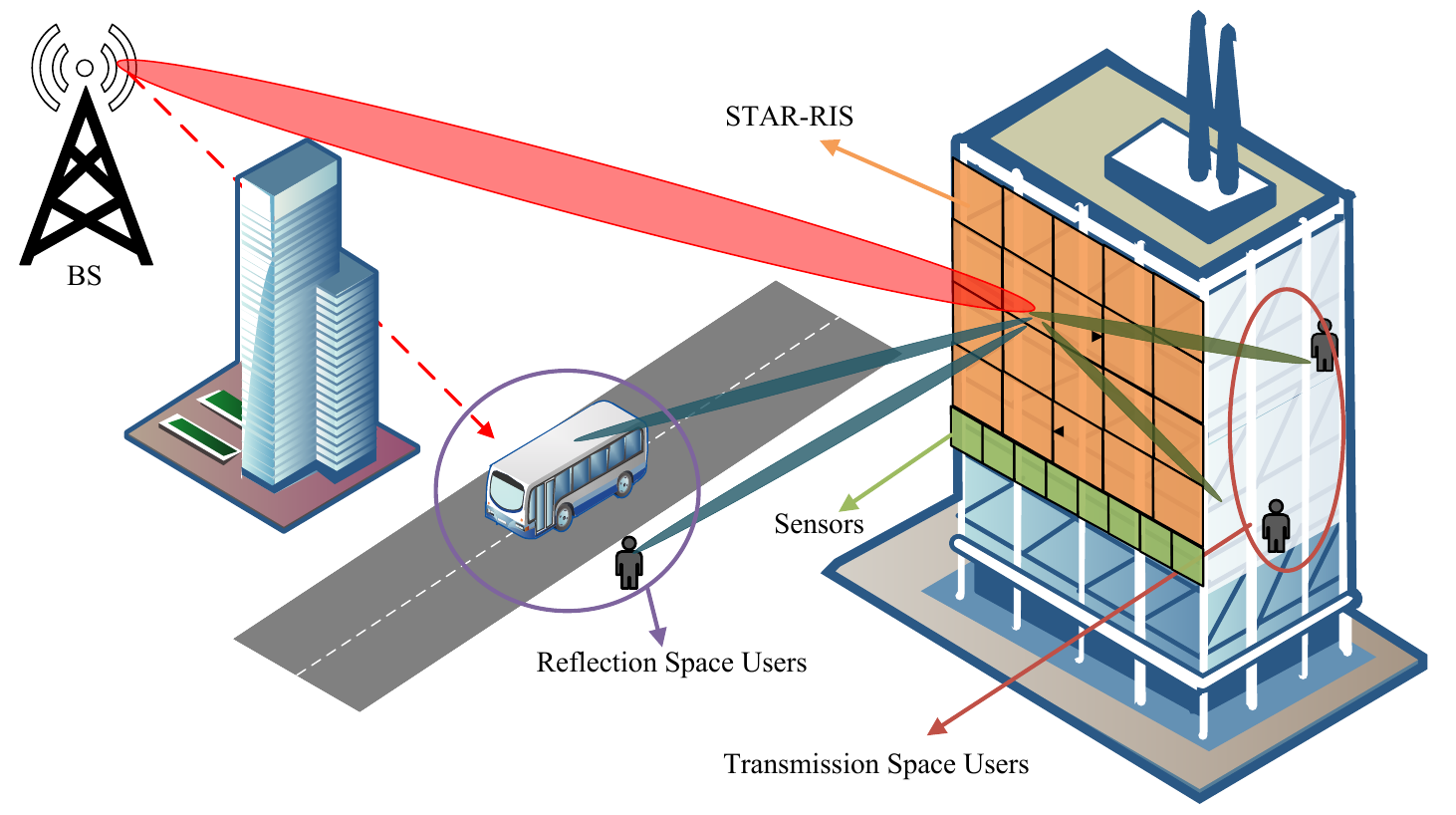}
  \caption{The considered STAR-RIS-aided ISAC system.}
    \label{fig:fig_1}
\end{figure}

\section{System model and basic assumptions}
\label{sec:system}

As shown in Fig.~\ref{fig:fig_1}, we consider an 
ISAC system, where sensing and communication tasks 
are assisted by a STAR-RIS placed on the outer wall 
of a smart building, which is a realistic deployment 
strategy in future smart cities and 6G intelligent 
environments \cite{Ahmed.2023}. Smart buildings 
already integrate multiple technologies (e.g., 
heating, ventilation, air conditioning, lighting, 
and surveillance). In accordance with the emerging 
paradigm of user-centric and environment-aware 
networks, we envision a smart building additionally 
incorporating radio-frequency (RF) sensors aimed at 
detecting and localizing low-mobility users in the 
vicinity of the facade \cite{Mehmood.2017.CommMag}. 
Sensors on the wall have direct visibility toward 
the front zone of the building, making them ideal 
for real-time, accurate localization, which is a 
crucial issue in many applications of environmental 
sensing, such as crowd monitoring, pedestrian flow 
analysis, public transportation systems monitoring 
\cite{Darsena.2023}, and traffic control.

The considered ISAC system consists of a BS equipped 
with a uniform linear array (ULA) of $M$ antennas, 
which transmits in downlink to $K$ users. The 
STAR-RIS is composed of a uniform planar array (UPA) 
of $N = N_x \times N_z$ reconfigurable elements, 
which enable full-space control. The impinging EM 
signals can be reflected from the STAR-RIS for 
outdoor sensing and communication, and can be 
simultaneously transmitted (i.e., refracted) for 
indoor communication.

We assume that $K_\text{R}$ single-antenna users 
reside in the outdoor reflection side ({\em 
reflection space}) and $K_\text{T}$ single-antenna 
users lie in the indoor transmission side ({\em 
transmission space}). Hereinafter, for the sake of 
conciseness, the set $\mathcal{K} \eqdef 
\{1,2,\ldots, K\}$, with $K \eqdef 
K_\text{T}+K_\text{R}$, collects the indexes of 
both indoor and outdoor receivers: the first 
$K_\text{T}$ entries of $\mathcal{K}$ (belonging 
to the set $\mathcal{K}_\text{T} \eqdef 
\{1,2,\ldots, K_\text{T}\}$) identify the indoor 
terminals, whereas the remaining $K_\text{R}$ ones 
(belonging to the set $\mathcal{K}_\text{R} \eqdef 
\{K_\text{T}+1,K_\text{T}+2,\ldots, K\}$) are used 
to index the outdoor ones. Similarly 
to~\cite{Wang.2023.TWC}, we consider a 
sensing-at-STAR-RIS structure, where a dedicated 
low-cost sensor, composed by a ULA with of $N_\text{s}$ 
elements, is mounted on the STAR-RIS. The sensor 
observations are conveyed to the BS via a dedicated 
low-latency control link, which is assumed to 
introduce negligible delay relative to the slot 
duration $T$ (the impact of feedback quantization 
errors on system performance is left for future 
investigation).

The direct BS-to-user links are assumed negligible 
in received power compared to the STAR-RIS-assisted 
paths, due to severe blockage and unfavorable 
propagation conditions typical of indoor and outdoor 
urban environments. Such an assumption is supported 
by standardized models and field measurements at 
millimeter-wave (mmWave) and sub-terahertz (sub-THz) 
frequencies, which show that building penetration 
loss and blockage effects may render direct paths 
negligible. For example, 3GPP TR~38.901 
\cite{3GPP_TR38.901} and ITU-R P.2109 
\cite{ITU-RP.2109-2} report penetration losses often 
exceeding $30$--$60$~dB for common construction 
materials, while extensive urban measurements (e.g., 
NYU Wireless \cite{Rapp_2015}) confirm that severe 
attenuation due to buildings or foliage is not rare 
in dense urban outdoor deployments.

The observation interval is divided into time slots 
of duration $T$ over which the channels and DoA 
parameters of the users are assumed to be constant. 

All the relevant channels are modeled as 
frequency-flat (i.e., the channel delay spread 
is small relative to the inverse signal 
bandwidth), as detailed in Subsection~II-\ref{sec:channel_model}. 
This assumption is well justified in narrowband 
deployments. In wideband scenarios, frequency 
selectivity can be handled by adopting an 
orthogonal frequency-division multiplexing 
(OFDM) waveform and applying the proposed 
framework independently to each subcarrier, 
since the channel response is approximately 
flat over the subcarrier bandwidth 
\cite{Liu.2020.TCOM}. 
During the $t$-th slot, with $t \in \mathbb{N}_0$, 
the DoA of the $k$-th user is described by 
$\db_k(t) \eqdef [\phi_k(t),\varphi_k(t)]^\trasp 
\in \mathbb{R}^2$, where $\phi_k(t)$ and 
$\varphi_k(t)$ are the elevation and azimuth DoAs 
between the $k$-th user and the STAR-RIS, 
respectively, for $k \in \mathcal{K}$.

We assume perfectly known DoAs for indoor 
users \cite{Zafari.2019}, i.e., $\db_1(t), 
\db_2(t), \ldots, \db_{K_\text{T}}(t)$ are treated 
as known deterministic parameters in the optimization 
process. 
Such an assumption is justified by the fact that 
indoor users are typically stationary or 
quasi-stationary, residing in fixed positions 
within a well-defined indoor environment. 
Consequently, their DoAs relative to the STAR-RIS 
can be accurately estimated during an initial 
calibration phase and reliably reused across 
multiple slots without significant performance 
degradation.
Conversely, the DoAs of 
outdoor users cannot be assumed perfectly known, 
since outdoor environments are typically dynamic and 
unpredictable, making DoAs of outdoor users 
observable only through estimation mechanisms 
working on a slot-by-slot basis 
\cite{Ziming_DoA.ArXiv}.

\begin{figure}[t]
    \centering
    \includegraphics[width=1\linewidth]{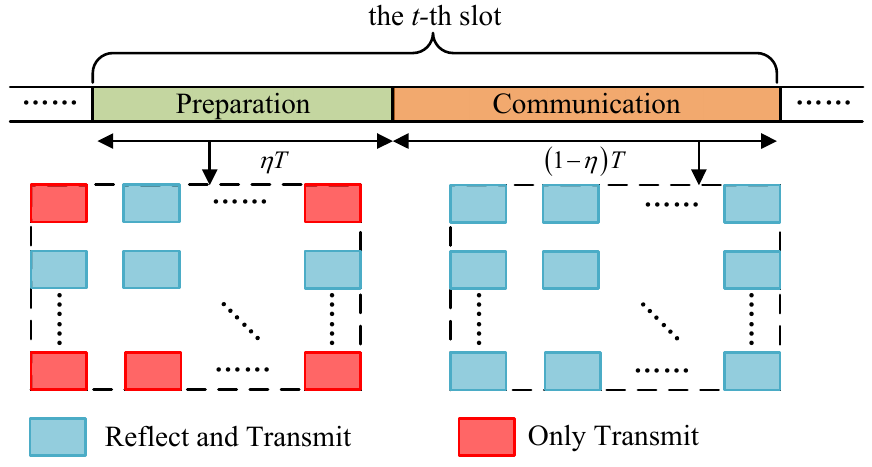}
    \caption{Working modes of the STAR-RIS in a generic slot.}
    \label{fig:fig_2}
\end{figure}

As depicted in Fig.~\ref{fig:fig_2}, each time slot $t$ 
is divided into two subslots. The first one, of 
duration $\eta \, T$ 
(with $0 \le \eta \le 1$ representing a system parameter) 
is referred to as the 
{\em preparation phase} or {\em stage}, and is 
dedicated to sensing of the outdoor users, i.e., 
estimation of the DoAs 
$\db_{K_\text{T}+1}(t), \db_{K_\text{T}+2}(t), 
\ldots, \db_{K}(t)$, and downlink communication to 
both indoor and outdoor users. During such a phase, 
reflection signals towards the outdoor users are 
realized by the STAR-RIS relying on the estimates 
of the outdoor user DoAs obtained in the 
previous $(t-1)$-th slot (see 
Subsections~\ref{sec:channel_model} 
and~\ref{sec:sensing}). Capitalizing on the 
estimates $\widehat{\db}_{K_\text{T}+1}(t), 
\widehat{\db}_{K_\text{T}+2}(t), \ldots, 
\widehat{\db}_{K}(t)$ of the outdoor user 
DoAs obtained at the end of the preparation 
stage during the $t$-th slot, the remaining part 
of the slot of duration $(1-\eta) \, T$, referred 
to as {\em communication phase/stage}, is only 
devoted to downlink communication 
to all the users.

As the system operates on a slot-by-slot basis, 
the analysis is restricted, without loss of 
generality, to a generic time slot. Accordingly, 
in the following the explicit dependence on the 
slot index $t$ is omitted, unless otherwise 
indicated.

\subsection{Signal transmitted by the BS}
\label{sec:tx-signal}

We assume that sensing of the outdoor devices is 
performed using the same signals employed for 
communication. 
During a given time slot, the signal transmitted 
by the BS to the $k$-th user in the $i$-th symbol 
interval can be written as
\begin{multline}
\bm{x}(i) = \bm{W} \, \bm{s}(i) = 
\sum_{k = 1}^{K} \bm{w}_k \, s_k(i)\:,  
\\ 
\text{for } i \in \mathcal{I} \eqdef 
\{0,1,\ldots, I-1\}
\label{eq:BS-transmitted}
\end{multline}
where $I$ is the number of symbols packed by the 
BS in a slot, $\bm{W} = [\bm{w}_1, \bm{w}_2, 
\ldots, \bm{w}_K] \in \mathbb{C}^{M \times K}$ 
represents the beamforming matrix, and 
$\bm{s}(i) = [s_1(i), s_2(i), \ldots, 
s_K(i)]^\mathsf{T} \in \mathbb{C}^{K}$ denotes 
the signal vector, with ${s_k}(i)$ 
($k \in \mathcal{K}$) denoting the signal destined 
to the $k$-th user. In the sequel, we assume that 
$\bm{s}(i)$ can be modeled as a zero-mean complex 
circular vector having covariance matrix 
$\mathbb{E}[\bm{s}(i) \, \bm{s}^{\mathsf{H}}(i)] 
= \mathbf{I}_{K}$, which is statistically 
independent of the information-bearing symbols 
transmitted in other symbol intervals and slots. 
This signal model remains valid across all 
operational modes of the system. Using 
\eqref{eq:BS-transmitted}, the total power 
transmitted by the BS is given by
\begin{equation}
\mathcal{P} \eqdef \mathbb{E}[\|\bm{x}(i)\|^2] =
\mathrm{tr}(\bm{W} \, \bm{W}^\herm) \:.
\label{eq:Prad}
\end{equation}

\subsection{STAR-RIS model}
\label{sec:RIS}

We assume that the STAR-RIS is located in the 
far-field region of the BS and, additionally, 
all the users are in the far-field region of the 
STAR-RIS. Moreover, we make the customary 
assumption that the one-sided bandwidth $B$ of 
the ISAC signal is much smaller than its carrier 
frequency $f_0$, thereby ensuring that the 
responses of the BS array and the RIS are essentially 
constant within the frequency interval 
$(f_0-\tfrac{B}{2},f_0+\tfrac{B}{2})$. Finally, 
we consider a linearly polarized (horizontally or 
vertically) wireless signal, and we neglect 
cross-polarization effects and losses. The RIS 
elements, which are arranged in a rectangular 
grid, are indexed row by row.
The adopted STAR-RIS model describes each element 
through its complex transmission and reflection 
coefficients, subject to energy conservation and 
phase-coupling constraints \cite{Mu.2022.TWC,
Xu.2022}. 
While more sophisticated models 
incorporating, e.g., mutual coupling, hardware 
impairments, or DoA-dependent responses have 
been proposed in the literature, the adopted 
model captures the essential physics of passive 
lossless STAR-RIS elements and is widely used 
as the reference framework in the STAR-RIS 
optimization literature 
\cite{Wang.2023.TWC,Zhang.2024.TWC,Zhang.2025}. 
Extensions to more general EM models 
are left for future work.

If the elements are designed to ensure 
weak dependence on the incidence DoA 
\cite{Shap.2022}, the EM response of the $n$-th 
element during the preparation stage is described 
by the transmission and reflection coefficients
\begin{equation}
\phi_{\text{T},n}^\text{p} = 
\beta_{\text{T},n}^\text{p} \, 
e^{j\theta_{\text{T},n}^\text{p}}
\label{eq:phi-pT}
\end{equation}
\begin{equation}
\phi_{\text{R},n}^\text{p} = 
\beta_{\text{R},n}^\text{p} \, 
e^{j\theta_{\text{R},n}^\text{p}}
\label{eq:phi-pR}
\end{equation}
respectively, 
where $j\eqdef \sqrt{-1}$ denotes the imaginary 
unit. Here, $\beta_{\text{a},n}^\text{p}$ and $\theta_{\text{a},n}^\text{p}$ denote the amplitude and phase responses of the $n$-th element during the preparation stage, respectively, with $\text{a} \in \{\text{T},\text{R}\}$ and $n \in \mathcal{N} \eqdef \{1,2,\ldots,N\}$. The coefficients $\phi_{\text{T},n}^\text{p}$ and $\phi_{\text{R},n}^\text{p}$ determine the transmitted and reflected fractions of the incident wave, respectively. 
The coupling between the reflection and 
transmission coefficients of each element depends 
on the operational mode of the STAR-RIS.

During the preparation stage, each element of the 
STAR-RIS may independently operate either in ES 
or TO mode \cite{Verde.2024.SPM}. In the case of 
ES, the signal impinging on the $n$-th element 
of the STAR-RIS is simultaneously reflected 
toward the outdoor region and transmitted toward 
the indoor region. Two constraints must be 
satisfied for a passive lossless STAR-RIS element 
in ES mode. First, according to the law of energy 
conservation \cite{Zhou.2014}, the incident 
signal power must equal the sum of reflected and 
transmitted powers. Second, since the scalar 
electric and magnetic impedances of the $n$-th 
element should have purely imaginary values, the 
phases of the reflection and transmission 
coefficients obey 
$\cos(\theta_{\text{R},n}^\text{p} - 
\theta_{\text{T},n}^\text{p}) = 0$ 
\cite{Xu.2022}. Specifically, the two constraints 
in ES mode are
\begin{align}
\left(\beta_{\text{R},n}^\text{p}\right)^2 + 
\left(\beta_{\text{T},n}^\text{p}\right)^2 
&= 1  
\label{2} \\
\theta_{\text{R},n}^\text{p} - 
\theta_{\text{T},n}^\text{p}  
&= \tfrac{\pi}{2} \; \text{or} \; 
\tfrac{3\pi}{2} \:.
\label{3}
\end{align}
Alternatively, for the TO configuration, the 
$n$-th element operates during the preparation 
stage in transmission mode solely, i.e.,
\begin{align}
\beta_{\text{R},n}^\text{p} &= 0 \quad 
\text{and} \quad \beta_{\text{T},n}^\text{p}=1
\label{2-1} \\
\theta_{\text{T},n}^\text{p} &\in [0,2 \pi) \:.
\label{3-1}
\end{align}
Therefore, transmission and reflection 
coefficients are straightforwardly decoupled in 
TO mode.

To characterize the two operating 
configurations during the preparation stage in 
a compact manner, we introduce a binary selection 
vector $\bm{b} \eqdef [b_1, b_2, \ldots, 
b_N]^\trasp \in \{0,1\}^N$, where each 
entry $b_n \in \{0,1\}$ indicates the working 
mode of the $n$-th element of the STAR-RIS. 
Specifically, when $b_n=1$, the $n$-th element 
operates in ES mode and, thus, the transmission 
and reflection coefficients \eqref{eq:phi-pT} 
and \eqref{eq:phi-pR} obey the constraints 
\eqref{2} and \eqref{3}. Otherwise, 
$\phi_{\text{T},n}^\text{p}$ and 
$\phi_{\text{R},n}^\text{p}$ fulfill the 
constraints \eqref{2-1} and \eqref{3-1} if 
$b_n=0$, and hence the $n$-th element works in 
TO manner. By constraining the $\ell^1$-norm of 
the vector $\bm{b}$ to a specific value 
$N_\text{part} \le N$, one may enforce a 
{\em dynamic partition} of the STAR-RIS during 
the preparation phase, according to which 
$N_\text{part}$ out of its $N$ elements work 
in ES mode, while the remaining 
$N - N_\text{part}$ ones are TO elements. Such 
a partition allows the system to flexibly 
balance communication throughput and sensing 
accuracy in the preparation phase.

In the communication stage, {\em all} the 
elements of the STAR-RIS are optimized to 
maximize communication performance for both 
indoor and outdoor users, i.e.,
\begin{align}
\left(\beta_{\text{R},n}^\text{c}\right)^2 + 
\left(\beta_{\text{T},n}^\text{c}\right)^2 
&= 1  
\label{eq:phi-cT-2} \\
\theta_{\text{R},n}^\text{c} - 
\theta_{\text{T},n}^\text{c}  
&= \tfrac{\pi}{2} \; \text{or} \; 
\tfrac{3\pi}{2}
\label{eq:phi-cT-3}
\end{align}
$\forall n \in \{1,2,\ldots,N\}$, where 
$\beta_{\text{a},n}^\text{c}$ and 
$\theta_{\text{a},n}^\text{c}$ represent the 
amplitude and phase of the $n$-th element of 
the STAR-RIS during the communication stage, 
respectively, for $\text{a} \in 
\{\text{T},\text{R}\}$.

For $\text{v} \in \{\text{p},\text{c}\}$, the 
STAR-RIS is collectively described by the diagonal 
matrices $\bm{\Phi}_{\text{R}}^\text{v} \in 
\mathbb{C}^{N \times N}$ and 
$\bm{\Phi}_{\text{T}}^\text{v} \in 
\mathbb{C}^{N \times N}$, gathering the reflection 
and transmission coefficients of the STAR-RIS, 
respectively. During the communication stage, all 
elements operate in ES mode, and the matrices are 
\begin{equation}
\bm{\Phi}_\text{a}^\text{c} \eqdef 
\mathrm{diag}\!\left(\beta_{\text{a},1}^\text{c}
e^{j\theta_{\text{a},1}^\text{c}}, \ldots, 
\beta_{\text{a},N}^\text{c}
e^{j\theta_{\text{a},N}^\text{c}}\right)
\label{eq:TC-RC-matrix}
\end{equation}
for $\text{a} \in \{\text{T},\text{R}\}$.
During the preparation stage, the situation is 
more involved due to the mixed ES/TO partition. 
The transmission matrix retains the same structure,
\begin{equation}
\bm{\Phi}_\text{T}^\text{p} \eqdef 
\mathrm{diag}\!\left(\beta_{\text{T},1}^\text{p}
e^{j\theta_{\text{T},1}^\text{p}}, \ldots, 
\beta_{\text{T},N}^\text{p}
e^{j\theta_{\text{T},N}^\text{p}}\right).
\label{eq:TC-RC-matrix-T-p}
\end{equation}
For the reflection matrix, however, the binary 
selection vector $\bm{b}$ must be accounted for 
explicitly: TO elements ($b_n = 0$) have zero 
reflection amplitude by definition~\eqref{2-1}, 
while ES elements ($b_n = 1$) have a free 
reflection amplitude $\widetilde{\beta}_{\text{R},n}^\text{p} 
\in [0,1]$ subject to the energy conservation 
constraint~\eqref{2}. It is therefore convenient 
to factor out the binary selection as 
\begin{equation}
\bm{\Phi}_\text{R}^\text{p} \eqdef 
\mathbf{B}\,\widetilde{\bm{\Phi}}_\text{R}^\text{p}
= \mathbf{B}\,\mathrm{diag}\!\left(
\widetilde{\beta}_{\text{R},1}^\text{p}
e^{j\theta_{\text{R},1}^\text{p}}, \ldots, 
\widetilde{\beta}_{\text{R},N}^\text{p}
e^{j\theta_{\text{R},N}^\text{p}}\right)
\label{eq:TC-RC-matrix-R-p}
\end{equation}
with $\mathbf{B} \eqdef \mathrm{diag}(b_1, 
b_2, \ldots, b_N)$, so that the reflection 
amplitude of the $n$-th element during the 
preparation stage is 
$\beta_{\text{R},n}^\text{p} = b_n\,
\widetilde{\beta}_{\text{R},n}^\text{p}$, 
for $n \in \mathcal{N}$.
It is worthwhile to note 
that the proposed optimization framework 
involves not only the physical parameters of 
the STAR-RIS (i.e., amplitudes and phases) but 
also the mode-selection vector $\bm{b}$ in the 
preparation stage.

\begin{figure}[t]
    \centering
    \includegraphics[width=1\linewidth]{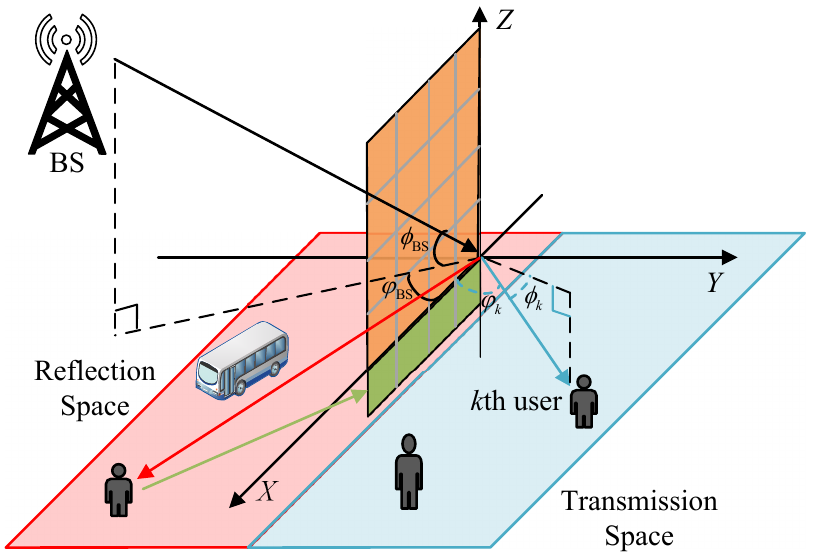}
    \caption{Angular representation of the ISAC system model.}
    \label{fig:3}
\end{figure}


\subsection{Channel model}
\label{sec:channel_model}

The two-stage protocol assumes that both the 
preparation and communication phases fit within 
a single coherence block of duration $T$, over 
which all channels are assumed to remain constant. 
This assumption is justified in scenarios with 
low-to-moderate user mobility, where the coherence 
time is sufficiently large to accommodate both 
phases. Under this condition, channel variations 
between the two stages are negligible by design.

For all wireless links, we employ a quasi-static 
frequency-flat Rician fading channel model 
\cite{Mu.2022.TWC}. Specifically, denoting with 
$\H_{\text{1}} \in \Cset^{N \times M}$ the channel 
from the BS to the STAR-RIS and with $\H_{\text{2}} 
\eqdef [\H_{\text{2,in}}^\trasp,
\H_{\text{2,out}}^\trasp]^\trasp \in \Cset^{K 
\times N}$ the channel from the STAR-RIS to the 
$K$ users, with $\H_{\text{2,in}} \in 
\Cset^{K_\text{T} \times N}$ and $\H_{\text{2,out}} 
\in \Cset^{K_\text{R} \times N}$, the composite 
channels are 
\begin{align}
\H_{\text{1}} &= 
\sqrt{\frac{1}{\varsigma_{\text{1}}}} 
\left( \sqrt{\frac{\mu_{\text{1}}}{1 + 
\mu_{\text{1}}}} \H_{\text{1}}^{\text{LoS}} + 
\sqrt{\frac{1}{1 + \mu_{\text{1}}}} 
\H_{\text{1}}^{\text{NLoS}} \right)
\label{eq:BS-channel} \\
\H_{\text{2,in}} &= 
\mathbf{P}_{\text{in}}^{-1/2} 
\left( \sqrt{\frac{\mu_{\text{2,in}}}{1 + 
\mu_{\text{2,in}}}} \H_{\text{2,in}}^{\text{LoS}} 
+ \sqrt{\frac{1}{1 + \mu_{\text{2,in}}}} 
\H_{\text{2,in}}^{\text{NLoS}} \right)
\label{eq:SU-channel-in} \\
\H_{\text{2,out}} &= 
\mathbf{P}_{\text{out}}^{-1/2} 
\left( \sqrt{\frac{\mu_{\text{2,out}}}{1 + 
\mu_{\text{2,out}}}} 
\H_{\text{2,out}}^{\text{LoS}} + 
\sqrt{\frac{1}{1 + \mu_{\text{2,out}}}} 
\H_{\text{2,out}}^{\text{NLoS}} \right)
\label{eq:SU-channel-out}
\end{align}
where $\varsigma_1$ is the path loss of the 
BS-to-STAR-RIS channel, whereas $\mathbf{P}_{\text{in}} 
\eqdef \mathrm{diag}(\varsigma_{\text{in},1},
\varsigma_{\text{in},2}, \ldots, 
\varsigma_{\text{in},K_\text{T}})$ collects the 
path losses of the channels between the STAR-RIS 
and the indoor users, and $\mathbf{P}_{\text{out}} 
\eqdef \mathrm{diag}(\varsigma_{\text{out},1},
\varsigma_{\text{out},2}, \ldots, 
\varsigma_{\text{out},K_\text{R}})$ gathers the 
path losses of the channels between the STAR-RIS 
and the outdoor users. 
For the generic $\ell$-th link, we adopt the 
path-loss model $\varsigma_\ell = \varsigma_0 
(d_\ell/d_0)^{-\kappa_\ell}$, with $d_\ell$ 
denoting the propagation distance, $d_0$ being 
the reference distance, $\varsigma_0$ the 
reference path loss at $d_0$, and $\kappa_\ell$ 
the path loss exponent. In principle, $d_0$ may 
differ between indoor and outdoor links --
typically $d_0 = 1$~m for indoor channels and 
$d_0 = 10$--$100$~m for outdoor channels 
\cite{Rapp_2015} -- however, since the product 
$\varsigma_0 \, d_0^{\kappa_\ell}$ is the 
physically meaningful quantity, a common 
reference distance $d_0 = 1$~m is adopted 
throughout, with $\varsigma_0$ and $\kappa_\ell$ 
calibrated accordingly for each link type.
Moreover, in 
\eqref{eq:BS-channel}--\eqref{eq:SU-channel-out}, 
$\mu_{1}, \mu_\text{2,in}, \mu_\text{2,out}>0$ 
are the Rice factors of the corresponding channels, 
whereas $\{\H_{\text{2,in}}^{\text{LoS}},
\H_{\text{2,out}}^{\text{LoS}}\}$ and 
$\{\H_{\text{2,in}}^{\text{NLoS}},
\H_{\text{2,out}}^{\text{NLoS}}\}$ denote the 
(deterministic) LoS component and 
the (random) NLoS components,
respectively.

The NLoS channel entries of the matrices 
$\H_1^{\text{NLoS}}$ and $\H_2^{\text{NLoS}} 
\eqdef [\{\H_{\text{2,in}}^{\text{NLoS}}\}^\trasp,
\{\H_{\text{2,out}}^{\text{NLoS}}\}^\trasp]^\trasp 
\in \Cset^{K \times N}$ are mutually independent, 
modeled as i.i.d. complex circularly-symmetric 
zero-mean unit-variance Gaussian random variables. 
For the LoS components, we express them using the 
array steering vectors of the BS and STAR-RIS. 
Specifically, under the assumption of 
half-wavelength inter-element spacing, one has 
(see Fig.~\ref{fig:3})
\begin{equation}
\bm{a}_{\text{BS}} ( \phi, \varphi) \eqdef 
\left[ 1, e^{- j \pi \cos\phi \, \cos\varphi}, 
\cdots, e^{- j \pi (M-1) \cos\phi \, 
\cos\varphi} \right]^\trasp
\end{equation}
and
\begin{multline}
\bm{a}_{\text{STAR}} ( \phi ,\varphi ) 
= 
\left[ 1, e^{-j\pi \sin\phi \, \cos\varphi}, 
\right. \\ \left. \cdots , 
e^{-j\pi (N_x-1) \sin\phi \, \cos\varphi} 
\right]^\trasp 
\\ \otimes 
\left[ 1, e^{-j\pi \sin\varphi}, \cdots, 
e^{-j\pi (N_z-1) \sin\varphi} \right]^\trasp 
\in \Cset^N
\label{eq:astar}
\end{multline}
respectively, with $\phi$ and $\varphi$ denoting 
the elevation and azimuth DoAs. Consequently, 
it turns out that
\begin{equation}
\H_{\text{1}}^{\text{LoS}} = 
\bm{a}_{\text{STAR}} (\phi_{\text{BS}},
\varphi_{\text{BS}}) \, 
\bm{a}_{\text{BS}}^\herm(\phi_{\text{BS}},
\varphi_{\text{BS}})
\end{equation}
with $\phi_{\text{BS}}$ and $\varphi_{\text{BS}}$ 
(see Fig.~\ref{fig:3}) denoting the elevation and 
azimuth DoAs between the BS and the STAR-RIS. 
Moreover, the $k$-th row of $\H_2^{\text{LoS}} 
\eqdef [\{\H_{\text{2,in}}^{\text{LoS}}\}^\trasp,
\{\H_{\text{2,out}}^{\text{LoS}}\}^\trasp]^\trasp 
\in \Cset^{K \times N}$ is given by 
$\bm{a}_{\text{STAR}}^\herm(\phi_k,\varphi_k)$, 
where $\phi_k$ and $\varphi_k$ are the elevation 
and azimuth DoAs between the $k$-th user and 
the STAR-RIS.

The three channels in 
\eqref{eq:BS-channel}--\eqref{eq:SU-channel-out} 
are treated differently in the optimization 
framework, reflecting their distinct estimation 
characteristics in the considered deployment.
The BS-to-STAR-RIS channel $\H_1$ is assumed to 
be perfectly known. This assumption is justified 
by the quasi-static nature of the BS-STAR-RIS 
link, which involves fixed infrastructure 
components whose geometry changes negligibly over 
time. Consequently, $\H_1$ can be accurately 
estimated during an offline calibration phase 
and reliably reused across multiple slots without 
significant performance degradation.
Regarding the links between the STAR-RIS and the 
indoor users, the matrix $\H_{\text{2,in}}$ is 
assumed to be perfectly known. This is justified 
by the fact that the DoAs of indoor users 
are given and the rich yet wide-sense stationary 
NLoS multipath of indoor users can be reliably 
estimated and tracked with  training 
overhead negligible relative to the coherence time. 
Equivalently, for each $k \in \mathcal{K}_{\text{T}}$, 
the corresponding channel vector $\h_{\text{2},k}$ 
is treated as deterministic in the optimization 
process, and its second-order matrix is given by
\begin{equation}
\Rb_{\h_{\text{2},k}\h_{\text{2},k}}
\eqdef 
\h_{\text{2},k}\h_{\text{2},k}^{\herm},
\quad \text{for $k \in \mathcal{K}_{\text{T}}$}.
\label{eq:indoor-rank-one-cov}
\end{equation}

On the other hand, as discussed earlier, the 
DoAs of outdoor users relative to the STAR-RIS 
are generally unknown and must be estimated 
\cite{Ziming_DoA.ArXiv}. Therefore, the LoS 
components of the STAR-RIS-to-outdoor-user 
channels inherently depend on imperfect 
DoA information. Moreover, in practical 
outdoor environments, the instantaneous diffuse 
NLoS component is difficult to acquire and track 
with negligible pilot overhead. For this reason, 
instead of assuming perfect instantaneous NLoS 
CSI for outdoor users, we characterize the 
STAR-RIS-to-outdoor-user links through their 
second-order statistics.

Specifically, for $k \in \mathcal{K}_{\text{R}}$ 
and $\mathrm{v} \in \{\mathrm{p},\mathrm{c}\}$, 
the channel vector associated with the $k$-th 
outdoor user is statistically modeled as
\begin{multline}
\left\{\h_{\text{2},k}^{\mathrm{v}}\right\}^{\herm}
=
\sqrt{\frac{1}{\varsigma_{\text{out},k}}}
\left(
\sqrt{\frac{\mu_{\text{2,out}}}{1+\mu_{\text{2,out}}}}
\right. \\ \left. \cdot
\bm{a}_{\text{STAR}}^{\herm}
\!\left(
\phi_k^{\mathrm{v}}+\epsilon_{\phi_k}^{\mathrm{v}},
\varphi_k^{\mathrm{v}}+\epsilon_{\varphi_k}^{\mathrm{v}}
\right)
\right.
\\
\left.
+
\sqrt{\frac{1}{1+\mu_{\text{2,out}}}}
\,
\bm{q}_k^{\herm}
\right),
\quad \text{ for $k \in \mathcal{K}_{\text{R}}$},
\label{eq:outdoor-statistical-channel}
\end{multline}
where $\bm{q}_k \in \Cset^{N}$ denotes the 
diffuse-scattering NLoS component of the $k$-th 
outdoor link. Different from the deterministic 
indoor CSI case, $\bm{q}_k$ is not required to 
be instantaneously known by the optimizer. 
Instead, it is described through its long-term 
spatial covariance $\Rb_{\bm{q},k} \eqdef 
\mathbb{E}[\bm{q}_k \bm{q}_k^{\herm}]$. This 
modeling approach captures the fact that the 
angular LoS component is mainly determined by 
the sensing process, whereas the diffuse 
scattering field can be learned and updated 
through its spatial statistics over a slower 
timescale.

Accordingly, the second-order channel matrix 
used for the statistical design is defined as
\begin{multline}
\Rb_{\h_{\text{2},k}\h_{\text{2},k}}^{\mathrm{v}}
\eqdef \mathbb{E}\!\left[ \h_{\text{2},k}^{\mathrm{v}} 
\left\{\h_{\text{2},k}^{\mathrm{v}}\right\}^{\herm} 
\right]
\\
= \frac{1}{\varsigma_{\text{out},k}} \left[ 
\frac{\mu_{\text{2,out}}}{1+\mu_{\text{2,out}}} 
\Rb_{\a,k}^{\mathrm{v}} + 
\frac{1}{1+\mu_{\text{2,out}}} \Rb_{\bm{q},k} 
\right.
\\
\left. + 
\frac{\sqrt{\mu_{\text{2,out}}}}{1+\mu_{\text{2,out}}}
\left( \Cb_{\a\bm{q},k}^{\mathrm{v}} + 
\left\{\Cb_{\a\bm{q},k}^{\mathrm{v}}\right\}^{\herm} 
\right) \right], 
\quad \text{for $k \in \mathcal{K}_{\text{R}}$},
\label{eq:outdoor-channel-covariance}
\end{multline}
where
\begin{multline}
\Rb_{\a,k}^{\mathrm{v}}
\eqdef
\mathbb{E}\!\left[
\bm{a}_{\text{STAR}}
\!\left(
\phi_k^{\mathrm{v}}+\epsilon_{\phi_k}^{\mathrm{v}},
\varphi_k^{\mathrm{v}}+\epsilon_{\varphi_k}^{\mathrm{v}}
\right)
\right.
\\
\left.
\times
\bm{a}_{\text{STAR}}^{\herm}
\!\left(
\phi_k^{\mathrm{v}}+\epsilon_{\phi_k}^{\mathrm{v}},
\varphi_k^{\mathrm{v}}+\epsilon_{\varphi_k}^{\mathrm{v}}
\right)
\right]
\label{eq:angular-steering-covariance}
\end{multline}
is the angular LoS covariance induced by the 
DoA estimation errors, and
\begin{equation}
\Cb_{\a\bm{q},k}^{\mathrm{v}}
\eqdef
\mathbb{E}\!\left[
\bm{a}_{\text{STAR}}
\!\left(
\phi_k^{\mathrm{v}}+\epsilon_{\phi_k}^{\mathrm{v}},
\varphi_k^{\mathrm{v}}+\epsilon_{\varphi_k}^{\mathrm{v}}
\right)
\bm{q}_k^{\herm}
\right]
\label{eq:los-nlos-cross-covariance}
\end{equation}
denotes the cross-covariance between the 
uncertain LoS steering component and the 
diffuse NLoS component.

The matrix $\Cb_{\a\bm{q},k}^{\mathrm{v}}$ is 
retained to describe possible statistical 
coupling between the dominant angular component 
and the diffuse scattering field. Such coupling 
may arise, for example, when dominant local 
scatterers are spatially related to the nominal 
LoS DoA, as in building facades, street 
canyons, or partially obstructed outdoor links. 
The conventional uncorrelated Rician model is 
recovered as a special case by setting 
$\Cb_{\a\bm{q},k}^{\mathrm{v}} = \mathbf{0}$. 
In the numerical evaluation, unless otherwise 
specified, this uncorrelated case is adopted as 
the default setting, while the impact of nonzero 
LoS--NLoS statistical coupling is investigated 
separately. It is worth emphasizing that the 
proposed optimization framework only requires 
the second-order matrices 
$\Rb_{\h_{\text{2},k}\h_{\text{2},k}}^{\mathrm{v}}$ 
for outdoor users, rather than the instantaneous 
realizations of their NLoS components. 
Instantaneous channel samples can still be 
generated for performance evaluation and 
benchmarking, but they are not assumed to be 
available for the beamforming and STAR-RIS 
coefficient optimization.

\subsection{Signal received by the $k$-th user}

According to Fig.~\ref{fig:fig_2}, we assume 
hereinafter that the number of symbols per slot 
is divided as $I = I_\text{p} + I_\text{c}$, 
where $I_\text{p}$ and $I_\text{c}$ are the 
number of symbols included into the preparation 
and communication phases, respectively. 
Consequently, it follows that $\eta = I_\text{p} 
\, T_\text{s}/T$, where $T_\text{s}$ denotes 
the symbol interval.

Let $\h_{\text{2},k}^\herm$ denote the $k$-th 
row of the matrix $\H_{\text{2}}$. During a 
given time slot, the signal received by the 
$k$-th user in the $i$-th symbol interval can 
be written as
\begin{multline}
y_{\text{u},k}(i) = 
\underbrace{\h_{\text{2},k}^\herm \, 
\bm{\Phi}_k(i) \, \H_{\text{1}} \, \w_k \, 
s_k(i)}_{\text{desired signal}}  
\\ +
\underbrace{\sum_{\shortstack{
\footnotesize $h=1$ \\ 
\footnotesize $h \neq k$}}^K 
\h_{\text{2},k}^\herm \, \bm{\Phi}_k(i) 
\, \H_{\text{1}} \, \w_h \, s_h(i) 
}_{\text{inter-user interference}} + 
n_{\text{u},k}(i)
\label{eq:RX-comm}
\end{multline}
for $k \in \mathcal{K}$ and $i \in 
\{0,1,\ldots,I-1\}$, where
\begin{equation}
\bm{\Phi}_k(i) = 
\begin{cases}
\bm{\Phi}^\text{p}(k) \:, & \text{for } 
i \in \{0,1, \ldots, I_\text{p}-1\} \\
\bm{\Phi}^\text{c}(k) \:, & \text{for } 
i \in \{I_\text{p},I_\text{p}+1, \ldots, I-1\}
\end{cases}
\label{eq:Phi-def}
\end{equation}
with
\begin{equation}
\bm{\Phi}^\text{v}(k) = 
\begin{cases}
\bm{\Phi}_\text{T}^\text{v} \:, & 
\text{for $k \in \mathcal{K}_\text{T}$} \\
\bm{\Phi}_\text{R}^\text{v} \:, & 
\text{for $k \in \mathcal{K}_\text{R}$} 
\end{cases}
\end{equation}
for $\text{v} \in \{\text{p},\text{c}\}$, and 
$n_{\text{u},k}(i) \sim \mathcal{CN}(0, 
\sigma_{n_\text{u}}^2)$ is complex circular additive white 
Gaussian noise (AWGN) at the $k$-th user 
terminal, with $n_{\text{u},k}(i_1)$ 
statistically independent of 
$n_{\text{u},k}(i_2)$ for $i_1 \neq i_2$.

Thus, when the $i$-th symbol belongs to stage 
$\mathrm{v} \in \{\mathrm{p},\mathrm{c}\}$, 
the instantaneous SINR of the $k$-th user can be written as
\begin{equation}
\gamma_{k}^\mathrm{v} 
= \frac{\left| \h_{\text{2},k}^{\herm} \, 
\bm{\Phi}^\mathrm{v}(k) \, \H_{\text{1}} \, 
\w_k \right|^2}
{\displaystyle 
\sum_{\shortstack{\footnotesize $h=1$ \\ 
\footnotesize $h \neq k$}}^K 
\left| \h_{\text{2},k}^{\herm} \, 
\bm{\Phi}^\mathrm{v}(k) \, \H_{\text{1}} \, 
\w_h \right|^2 + \sigma_{n_\text{u}}^2}
\label{9}
\end{equation}
for $k \in \mathcal{K}$ and $\mathrm{v} \in 
\{\mathrm{p},\mathrm{c}\}$. For indoor users, 
the channel vector $\h_{\text{2},k}$ is 
deterministic in the optimization process, as 
discussed in Subsection~II-\ref{sec:channel_model}. 
For outdoor users, however, the instantaneous 
channel realization is not assumed to be 
available for beamforming and STAR-RIS 
coefficient optimization. Therefore, the design 
is carried out by replacing the instantaneous 
quadratic channel terms in \eqref{9} with their 
corresponding second-order statistics.

To this end, we define the stage-dependent 
channel covariance matrix
\begin{equation}
\Rb^\mathrm{v}(k) \eqdef 
\begin{cases}
\h_{\text{2},k}\h_{\text{2},k}^{\herm}, 
& \text{for $k \in \mathcal{K}_{\mathrm{T}}$}\\[2mm]
\Rb_{\h_{\text{2},k}\h_{\text{2},k}}^\mathrm{v}, 
& \text{for $k \in \mathcal{K}_{\mathrm{R}}$}
\end{cases}
\label{eq:unified-channel-covariance}
\end{equation}
where 
$\Rb_{\h_{\text{2},k}\h_{\text{2},k}}^\mathrm{v}$ 
is given in Subsection~II-\ref{sec:channel_model} 
and accounts for both the DoA estimation 
uncertainty and the long-term diffuse-scattering 
outdoor statistics. Accordingly, the 
statistical SINR used for robust system design 
is given by
\begin{multline}
\!\!\!
\bar{\gamma}_{k}^\mathrm{v} 
= 
\frac{
\w_k^\herm \, \H_{\text{1}}^\herm  
\left\{\bm{\Phi}^\mathrm{v}(k)\right\}^*  
\Rb^\mathrm{v}(k) \, 
\bm{\Phi}^\mathrm{v}(k) \, \H_{\text{1}} \, \w_k
}
{
\displaystyle 
\sum_{\shortstack{\footnotesize $h=1$ \\ 
\footnotesize $h \neq k$}}^K \!\!
\w_h^\herm \, \H_{\text{1}}^\herm 
\left\{\bm{\Phi}^\mathrm{v}(k)\right\}^*  
\Rb^\mathrm{v}(k) \, 
\bm{\Phi}^\mathrm{v}(k) \, \H_{\text{1}} \, \w_h
+ \sigma_{n_\text{u}}^2
}
\label{eq:statistical-sinr}
\end{multline}
for $k \in \mathcal{K}$ and $\mathrm{v} \in 
\{\mathrm{p},\mathrm{c}\}$. It is noteworthy 
that \eqref{eq:statistical-sinr} coincides with 
the instantaneous SINR expression for indoor 
users, since their channel covariance reduces to 
the rank-one matrix 
$\h_{\text{2},k}\h_{\text{2},k}^{\herm}$. For 
outdoor users, instead, 
eq.~\eqref{eq:statistical-sinr} depends only on the 
second-order statistics, i.e.,  
on matrix 
$\Rb_{\h_{\text{2},k}\h_{\text{2},k}}^\mathrm{v}$.

Assuming that the BS encodes the information for 
each user using an i.i.d. Gaussian code, the 
achievable sum-rate or throughput 
of the system can be 
expressed as
\begin{equation}
\capa = \eta \, \capa^\text{p} + 
(1-\eta) \, \capa^\text{c}
\label{eq:rate}
\end{equation}
with
\begin{equation}
\capa^\text{v} = \sum_{k=1}^K 
\log_2\!\left(1 + \gamma_{k}^\text{v} \right)
\label{eq:rate-v}
\end{equation}
for $\text{v} \in \{\text{p},\text{c}\}$ 
representing the sum-rate in the 
preparation/communication stage. Since the 
instantaneous SINR $\gamma_k^\text{v}$ in 
\eqref{9} is not available for outdoor users, 
we replace it with the statistical SINR 
$\bar{\gamma}_k^\text{v}$ in 
\eqref{eq:statistical-sinr}, yielding the 
tractable statistical sum-rate
\begin{equation}
\bar{\capa} = \eta \, \bar{\capa}^\text{p} + 
(1-\eta) \, \bar{\capa}^\text{c}
\label{eq:bar-rate}
\end{equation}
with
\begin{equation}
\bar{\capa}^\text{v} = \sum_{k=1}^K 
\log_2\!\left(1 + \bar{\gamma}_{k}^\text{v} 
\right), \quad \text{v} \in \{\text{p},\text{c}\}.
\label{eq:bar-rate-v}
\end{equation}
In what follows, $\bar{\capa}$ is assumed as 
the overall communication performance metric 
used in the optimization.

\subsection{Signal received by the sensor}
\label{sec:sensing}

For analytical tractability, it is assumed that 
the sensor mounted on the STAR-RIS can perfectly 
separate the echo signals corresponding to the 
different targets (i.e., outdoor users). This is 
consistent with recent studies on ISAC systems 
\cite{Yuan.2021,Wang.2023.TWC,Zhang.2025}, which 
employ target-specific signaling or structured 
processing to make multi-target discrimination 
feasible. In particular, the echo signals can be 
processed independently if the targets are 
sufficiently separated in DoA, range, or 
Doppler domain \cite{Yuan.2021,Wang.2023.TWC}, 
or when orthogonal modulation sequences are 
employed \cite{Zhang.2025}. Such an assumption 
is widely adopted in the ISAC literature and can 
be relaxed in future work to incorporate partial 
or imperfect target separability.

During the preparation phase of a given time 
slot, under the aforementioned target-specific 
processing assumption, the received echo signal 
at the sensor associated with the $k$-th outdoor 
user can be written as
\begin{multline}
\bm{y}_{\text{s},k}(i) = 
\alpha_k \,
\bm{a}_{\text{S}}(\phi_k,\varphi_k) \,
\bm{a}_{\text{STAR}}^\herm(\phi_k,\varphi_k) \,
\bm{\Phi}_{\text{R}}^{\text{p}} \,
\H_{\text{1}} \,
\bm{x}(i)
\\
+ \bm{c}_{\text{s},k}(i)
+ \nb_{\text{s}}(i)
\label{eq:RX-sens}
\end{multline}
for $k \in \mathcal{K}_\text{R}$ and 
$i \in \{0,1,\ldots,I_\text{p}-1\}$. The first 
term in \eqref{eq:RX-sens} is the useful echo 
from the $k$-th outdoor user, whereas 
$\bm{c}_{\text{s},k}(i)$ denotes the residual 
clutter observed in the sensing cell of the 
$k$-th user after target-specific processing, 
and $\nb_{\text{s}}(i) \in \Cset^{N_\text{s}}$ 
is the sensing noise. The clutter term 
$\bm{c}_{\text{s},k}(i)$ models the aggregate 
contribution of static and slowly varying 
scatterers in the sensing cell; its long-term 
power is absorbed into the effective disturbance 
variance defined below, while short-term 
fluctuations are treated as part of the 
unmodeled disturbance.

The scalar $\alpha_k \in \Cset$ denotes the 
effective target response of the $k$-th outdoor 
user. It absorbs the round-trip propagation 
attenuation, the reflection loss, the radar 
cross section (RCS)-dependent scattering 
coefficient, and other slowly varying 
sensing-channel effects. These effects are 
represented through equivalent nominal 
parameters, which can be obtained from prior 
sensing, calibration, or long-term tracking. 
Possible target-response fluctuations are 
examined numerically by perturbing $\alpha_k$ 
around its nominal value.

The steering vector of the sensor with 
half-wavelength inter-element spacing is given 
by
\begin{multline}
\bm{a}_{\text{S}}({\phi_k},{\varphi_k}) =
\left[1, e^{-j\pi\cos{\phi_k}\cos{\varphi_k}},
\right.\\
\left.\cdots, 
e^{-j\pi(N_{\text{s}}-1)\cos{\phi_k}
\cos{\varphi_k}}\right]^\trasp .
\label{eq:sensor-steering}
\end{multline}
The sensing noise is modeled as zero-mean 
complex circular AWGN with covariance matrix 
$\mathbb{E}[\nb_{\text{s}}(i)
\nb_{\text{s}}^\herm(i)] = 
\sigma_{n_\text{s}}^2 \I_{N_\text{s}}$ and is 
assumed to be statistically independent across 
symbol intervals and time slots.

After target-specific processing, the residual 
clutter and unmodeled scattering components are 
absorbed into an effective sensing disturbance 
power. Accordingly, the sensing 
signal-to-noise ratio (SSNR) of the $k$-th 
outdoor user is defined as
\begin{equation}
\text{SSNR}_k =
\frac{|\alpha_k|^2
\left\|\bm{a}_{\text{S}}(\phi_k,\varphi_k)
\bm{a}_{\text{STAR}}^\herm(\phi_k,\varphi_k)
\bm{\Phi}_{\text{R}}^{\text{p}} 
\H_{\text{1}} \bm{W}\right\|^2}
{N_\text{s}\,\sigma_{\text{s},k,\text{eff}}^2}
\label{11}
\end{equation}
for $k \in \mathcal{K}_\text{R}$, where 
$\sigma_{\text{s},k,\text{eff}}^2$ denotes the 
effective disturbance variance, including thermal 
noise and the long-term residual clutter power 
in the sensing cell of the $k$-th outdoor user. 
This abstraction preserves the quadratic 
structure of the SSNR constraint with respect 
to the BS beamforming matrix and the STAR-RIS 
reflection coefficients, while the effects of 
RCS mismatch and residual clutter fluctuations 
can be evaluated by varying $\alpha_k$ and 
$\sigma_{\text{s},k,\text{eff}}^2$.

The metric \eqref{11} captures the impact of 
the BS beamformer, the STAR-RIS reflection 
coefficients, the target response, and the 
sensing noise level on the DoA-estimation 
quality. Although the SSNR does not explicitly 
include a clutter covariance in its denominator, 
it provides a tractable and widely used 
surrogate for sensing quality when the useful 
target echo is dominant after target-specific 
processing \cite{Liu.2020}.

Let $\bm{Y}_{\text{s},k} \eqdef 
[\bm{y}_{\text{s},k}(0), 
\bm{y}_{\text{s},k}(1), \ldots, 
\bm{y}_{\text{s},k}(I_\text{p}-1)] \in 
\Cset^{N_\text{s} \times I_\text{p}}$ collect 
the echo observations associated with the $k$-th 
outdoor user during the preparation phase. On 
the basis of $\bm{Y}_{\text{s},k}$, the task of 
the sensor is to provide an estimate 
$\widehat{\db}_k = [\widehat{\phi}_k,
\widehat{\varphi}_k]^\trasp \in \mathbb{R}^2$ 
of the $k$-th outdoor DoA during the 
current slot, which can be obtained by 
classic maximum likelihood estimation (MLE) 
\cite{Wang.2023.TWC,Kay} or subspace-based 
algorithms \cite{Stoica.1989,Roy.1989,
Shao.2022}. The mean square errors (MSEs) 
$\sigma^2_{\phi_k} \eqdef 
\mathbb{E}[(\phi_k-\widehat{\phi}_k)^2]$ and 
$\sigma^2_{\varphi_k} \eqdef 
\mathbb{E}[(\varphi_k-\widehat{\varphi}_k)^2]$ 
are commonly used to evaluate the estimation 
performance.

The posterior Cramér--Rao bound (PCRB) 
\cite{Van-Trees} establishes a lower bound on 
$\sigma^2_{\phi_k}$ and $\sigma^2_{\varphi_k}$ 
and, hence, it is a rigorous benchmark for 
DoA estimation accuracy. However, in 
STAR-RIS-enabled ISAC systems, the PCRB 
typically leads to highly intricate mathematical 
expressions \cite{Wang.2023.TWC}. Therefore, 
using the PCRB as performance metric 
significantly complicates the optimization 
problem and limits analytical tractability. To 
maintain a good balance between accuracy and 
computational simplicity, we adopt the SSNR as 
the primary performance indicator for the 
estimation of outdoor users' DoAs. This 
approach allows us to formulate the problem in 
a more tractable manner, while still capturing 
the essential impact of beamforming, propagation 
conditions, and STAR-RIS configuration on the 
achievable sensing performance. However, 
replacing the PCRB with the SSNR is meaningful 
primarily in the high-SSNR regime \cite{Van-Trees}. 
For this reason, we impose the accumulated sensing constraint 
$\eta \, \mathbb{E}[\mathrm{SSNR}_k] \ge \delta_{\text{sens}}$, 
with $\delta_\text{sens}>0$ being a given 
threshold, which ensures operation in the 
high-SSNR region and makes the SSNR-based 
uncertainty model a reliable approximation.

\section{Optimization framework}
\label{sec:optmization}

We formulate an optimization problem aimed at 
maximizing the achievable communication rate 
with respect to $\bm{W}$, 
$\widetilde{\bm{\Phi}}_\text{R}^\text{p}$ 
(encompassing $\bm{b}$ too), 
$\bm{\Phi}_\text{T}^\text{p}$, 
$\bm{\Phi}_\text{R}^\text{c}$, and 
$\bm{\Phi}_\text{T}^\text{c}$, while ensuring 
that both the BS transmit power constraint and 
the target detection performance requirement 
are satisfied. The considered optimization also 
takes into account the intrinsic physical 
limitations of the STAR-RIS elements, such as 
energy conservation, amplitude--phase coupling, 
and power-splitting constraints between 
transmission and reflection modes.

According to \eqref{eq:TC-RC-matrix-R-p}, the 
matrix $\bm{\Phi}_\text{R}^\text{p}$ depends on 
both the metasurface partition vector $\bm{b}$ 
and the STAR-RIS reflection coefficients during 
the preparation phase. For optimization purposes, 
we exploit the factorization 
$\bm{\Phi}_\text{R}^\text{p} = \mathbf{B} \, 
\widetilde{\bm{\Phi}}_\text{R}^\text{p}$, where 
$\mathbf{B} = \mathrm{diag}(\bm{b})$ encodes the 
metasurface partition, while
$\widetilde{\bm{\Phi}}_\text{R}^\text{p} \eqdef 
\mathrm{diag}(\widetilde{\beta}_{\text{R},1}^\text{p}
\, e^{j\theta_{\text{R},1}^\text{p}}, \ldots, 
\widetilde{\beta}_{\text{R},N}^\text{p} \, 
e^{j\theta_{\text{R},N}^\text{p}})$ contains 
only the reflection amplitudes and phases of 
the STAR-RIS during the preparation stage.

For $k \in \mathcal{K}_\text{R}$ and 
$\text{v} \in \{\text{p},\text{c}\}$, the SINR 
in \eqref{9} depends on the DoA $\db_k$ of the 
outdoor users, which must therefore be 
incorporated into the joint beamforming and 
STAR-RIS optimization. Since outdoor DoAs vary 
slowly relative to the slot duration $T$ 
(see Subsection~\ref{sec:tx-signal}), their estimates obtained 
in one slot can be reliably reused for system 
design in the preparation phase of the 
subsequent slot.

Hereinafter, for $k \in \mathcal{K}_\text{R}$, 
we denote with $\phi_k^\text{p}$ and 
$\varphi_k^\text{p}$ the DoAs of the outdoor 
users used to optimize the system during the 
preparation phase: they correspond to the 
DoA estimates of the outdoor users 
obtained in the previous slot. On the other 
hand, for $k \in \mathcal{K}_\text{R}$, the 
system design during the communication phase 
relies on the outdoor user DoAs 
$\phi_k^\text{c}$ and $\varphi_k^\text{c}$, 
estimated in the preparation stage of the same 
slot. Specifically, with reference to the 
$t$-th slot, we set 
$\phi_k^\text{p} = \widehat{\phi}_k(t-1)$ and 
$\varphi_k^\text{p} = \widehat{\varphi}_k(t-1)$ 
in the preparation phase, and 
$\phi_k^\text{c} = \widehat{\phi}_k(t)$ and 
$\varphi_k^\text{c} = \widehat{\varphi}_k(t)$ 
in the communication stage.

To enhance robustness against estimation errors 
of the outdoor users' DoAs, we average 
the achievable rate and the SSNR with respect 
to the random variables 
$\epsilon_{\phi_k}^\text{v} \eqdef 
\phi_k - \phi_k^\text{v}$ and 
$\epsilon_{\varphi_k}^\text{v} \eqdef 
\varphi_k - \varphi_k^\text{v}$, for 
$k \in \mathcal{K}_\text{R}$ and 
$\text{v} \in \{\text{p},\text{c}\}$. The 
optimization problem can thus be formulated as
\begin{subequations}
\begin{align}
\textbf{(P1)} \quad & 
\max_{\bm{W}, \widetilde{\bm{\Phi}}_\text{R}^\text{p}, 
\bm{b}, \bm{\Phi}_\text{T}^\text{p}, 
\bm{\Phi}_\text{R}^\text{c}, 
\bm{\Phi}_\text{T}^\text{c}}
\mathbb{E}[\capa] \label{12} \\
\text{s.t.} \quad 
& \phi_k = \phi_k^\text{p} + 
\epsilon_{\phi_k}^\text{p}, \;
\varphi_k = \varphi_k^\text{p} + 
\epsilon_{\varphi_k}^\text{p} \notag \\ 
& \text{for } k \in \mathcal{K}_\text{R} 
\text{ and } i \in \{0,\ldots,I_\text{p}-1\} 
\label{12ini-p} \\
& \phi_k = \phi_k^\text{c} + 
\epsilon_{\phi_k}^\text{c}, \;
\varphi_k = \varphi_k^\text{c} + 
\epsilon_{\varphi_k}^\text{c} \notag \\ 
& \text{for } k \in \mathcal{K}_\text{R} 
\text{ and } i \in 
\{I_\text{p},\ldots,I-1\} 
\label{12ini-c} \\
& \theta_{\text{R},n}^\text{v}, 
\theta_{\text{T},n}^\text{v} \in [0,2\pi), \;
\cos(\theta_{\text{R},n}^\text{v} - 
\theta_{\text{T},n}^\text{v}) = 0 \notag \\
& \beta_{\text{R},n}^\text{v},
\beta_{\text{T},n}^\text{v} \in [0,1], \;
(\beta_{\text{R},n}^\text{v})^2 + 
(\beta_{\text{T},n}^\text{v})^2 = 1 \notag \\
& \text{for } n \in \mathcal{N} 
\text{ and } \text{v} \in 
\{\text{p},\text{c}\} \label{12a} \\
& \eta \, \mathbb{E}[\mathrm{SSNR}(\db_k)] \ge 
\delta_\text{sens}, \; 
k \in \mathcal{K}_\text{R} \label{12b} \\
& \mathcal{P} \le \mathcal{P}_{\text{max}} 
\label{12c} \\
& \|\bm{b}\|_1 = N_\text{part} \le N 
\label{12d}
\end{align}
\end{subequations}
where \eqref{12ini-p} and \eqref{12ini-c} 
account for the fact that the optimization 
relies on the estimates of the outdoor users' 
DoAs, with the expectation taken with 
respect to the estimation errors 
$\epsilon_{\phi_k}^\text{v}$ and 
$\epsilon_{\varphi_k}^\text{v}$, for 
$k \in \mathcal{K}_\text{R}$ and 
$\text{v} \in \{\text{p},\text{c}\}$. 
Constraint \eqref{12a} ensures compliance 
with the physical operating conditions of the 
STAR-RIS; constraint \eqref{12b} guarantees 
that the accumulated average SSNR (ASSNR) of the targets 
remains above a given threshold 
$\delta_\text{sens}$ and is active only during 
the preparation phase; constraint \eqref{12c} 
ensures that the BS transmission power does 
not exceed its maximum limit 
$\mathcal{P}_{\text{max}} > 0$; and constraint 
\eqref{12d} partitions the STAR-RIS during the 
preparation phase, with $N_\text{part}$ 
elements operating in ES mode and the others 
in TO mode.

\subsection{Evaluation of the average 
performance metrics}

To calculate the average achievable rate and 
the average sensing SNR, we assume that the 
random pairs $(\epsilon_{\phi_{K_\text{T}+1}}^\text{v}, 
\epsilon_{\varphi_{K_\text{T}+1}}^\text{v}), 
\ldots, (\epsilon_{\phi_{K}}^\text{v}, 
\epsilon_{\varphi_{K}}^\text{v})$ are 
statistically independent, with 
$\text{v} \in \{\text{p},\text{c}\}$. For 
high SNR and large samples, DoA estimates 
from MLE and subspace-based algorithms are 
asymptotically normal with variance 
approaching the PCRB 
\cite{Van-Trees,Krim.1996}. Therefore, for 
each $k \in \mathcal{K}_\text{R}$, the 
estimation errors $\epsilon_{\phi_k}^\text{v}$ 
and $\epsilon_{\varphi_k}^\text{v}$ are 
modeled as independent zero-mean Gaussian 
random variables with variances 
$\sigma^2_{\phi_k}$ and $\sigma^2_{\varphi_k}$ 
given by the PCRB, whose 
expression can be found in 
\cite{Wang.2023.TWC}.

It is worth noting that the estimates 
$\phi_k^\text{p}$ and $\varphi_k^\text{p}$ 
used in the current slot were obtained during 
the previous slot. Consequently, they are 
treated as deterministic parameters in the 
optimization performed during the preparation 
stage. Likewise, $\phi_k^\text{c}$ and 
$\varphi_k^\text{c}$ have been acquired in 
the preparation phase of the current slot and 
are therefore regarded as deterministic 
quantities in the subsequent 
communication-stage optimization. 
Consequently, the DoAs $\phi_k$ and 
$\varphi_k$ of the outdoor users in 
\eqref{12ini-p} and \eqref{12ini-c} are 
independent Gaussian random variables with 
mean $\phi_k^\text{v}$ and 
$\varphi_k^\text{v}$, and variance 
$\sigma^2_{\phi_k}$ and $\sigma^2_{\varphi_k}$, 
respectively, for all 
$k \in \mathcal{K}_\text{R}$ and 
$\text{v} \in \{\text{p},\text{c}\}$.

It follows from \eqref{eq:rate} that
\begin{equation}
\mathbb{E}[\capa] = \eta \, 
\mathbb{E}[\capa^\text{p}] + (1-\eta) \, 
\mathbb{E}[\capa^\text{c}].
\label{eq:averate}
\end{equation}
The exact evaluation of 
$\mathbb{E}[\capa^\text{v}]$ is generally 
intractable, since the instantaneous SINR 
involves random angular errors and, for 
outdoor users, the instantaneous diffuse NLoS 
components are not assumed to be available in 
the proposed design. Therefore, we construct 
a tractable statistical surrogate by 
approximating $\mathbb{E}[\capa^\text{v}]$ 
with the statistical sum-rate $\bar{\capa}^\text{v}$ 
defined in \eqref{eq:bar-rate-v}, which is 
obtained by replacing the instantaneous SINR 
with the statistical SINR $\bar{\gamma}_k^\text{v}$ 
in \eqref{eq:statistical-sinr}. This 
approximation is equivalent to exchanging the 
expectation and the logarithm, which by 
Jensen's inequality yields an upper bound on 
$\mathbb{E}[\capa^\text{v}]$. The gap between 
the bound and the true expectation is 
negligible when the fluctuations of the SINR 
around its mean are small relative to the mean 
itself, a condition that is well satisfied in 
the moderate-to-high SINR regime targeted by 
the sensing constraint \eqref{12b}. 
Accordingly, problem \textbf{(P1)} is replaced 
by its tractable surrogate
\begin{equation}
\textbf{(P1-mod)} \quad
\max_{\bm{W}, \widetilde{\bm{\Phi}}_\text{R}^\text{p}, 
\bm{b}, \bm{\Phi}_\text{T}^\text{p}, 
\bm{\Phi}_\text{R}^\text{c}, 
\bm{\Phi}_\text{T}^\text{c}} \;
\bar{\capa}
\label{eq:P1mod}
\end{equation}
subject to the same constraints 
\eqref{12ini-p}--\eqref{12d}.

The ASSNR is obtained by averaging \eqref{11} 
with respect to $\epsilon_{\phi_k}^\text{p}$ 
and $\epsilon_{\varphi_k}^\text{p}$, thus 
yielding
\begin{multline}
\mathbb{E}[\mathrm{SSNR}(\db_k)] = 
\frac{ |\alpha_k|^2}
{N_\text{s} \, \sigma_{\text{s},k,\text{eff}}^2} 
\, \mathbb{E}\!\left[\left\| 
\bm{a}_{\text{S}}(\phi_k^\text{p} + 
\epsilon_{\phi_k}^\text{p}, \varphi_k^\text{p} + 
\epsilon_{\varphi_k}^\text{p}) \right.\right. 
\\ \left.\left. \cdot \,
\bm{a}_{\text{STAR}}^\herm(\phi_k^\text{p} + 
\epsilon_{\phi_k}^\text{p}, \varphi_k^\text{p} + 
\epsilon_{\varphi_k}^\text{p}) \, 
\bm{\Phi}_{\text{R}}^{\text{p}} \, 
\H_{\text{1}} \, \bm{W} 
\right\|^2\right]
\\ = 
\frac{|\alpha_k|^2}
{N_\text{s} \, \sigma_{\text{s},k,\text{eff}}^2} \,
\mathrm{tr}\!\left[ 
\bm{W}^\herm \, \H_{\text{1}}^\herm \, 
\{\bm{\Phi}_{\text{R}}^{\text{p}}\}^* \,
\Rb_{\bm{a}_{\text{S}} 
\bm{a}_{\text{STAR}},k}^\text{p} \, 
\bm{\Phi}_{\text{R}}^{\text{p}} \, 
\H_{\text{1}} \, \bm{W}
\right]
\label{ave-11}
\end{multline}
for $k \in \mathcal{K}_\text{R}$, with
\begin{multline}
\Rb_{\bm{a}_{\text{S}} 
\bm{a}_{\text{STAR}},k}^\text{p} \eqdef 
\mathbb{E}\!\left[\| 
\bm{a}_{\text{S}}(\phi_k^\text{p} + 
\epsilon_{\phi_k}^\text{p}, \varphi_k^\text{p} + 
\epsilon_{\varphi_k}^\text{p})\|^2 
\right. \\ \left. \cdot \,
\bm{a}_{\text{STAR}}(\phi_k^\text{p} + 
\epsilon_{\phi_k}^\text{p}, \varphi_k^\text{p} + 
\epsilon_{\varphi_k}^\text{p}) \,
\bm{a}_{\text{STAR}}^\herm(\phi_k^\text{p} + 
\epsilon_{\phi_k}^\text{p}, \varphi_k^\text{p} + 
\epsilon_{\varphi_k}^\text{p}) \right]
\end{multline}
under \eqref{12ini-p} and \eqref{12ini-c}. 
Similarly to 
$\Rb_{\h_{\text{2},k}\h_{\text{2},k}}^\text{v}$, 
the matrix 
$\Rb_{\bm{a}_{\text{S}} 
\bm{a}_{\text{STAR}},k}^\text{p}$ cannot be 
expressed in closed form and is therefore 
evaluated numerically in 
Section~\ref{sec:Simulation}.

\subsection{Closed-form fractional programming solution}

In the sequel, we develop an efficient solution 
to problem \textbf{(P1-mod)} defined in 
\eqref{eq:P1mod}, whose objective is the 
tractable statistical sum-rate $\bar{\capa}$ 
in \eqref{eq:bar-rate}, with $\bar{\gamma}_k^\text{v}$ 
given by \eqref{eq:statistical-sinr} and 
$\Rb^\text{v}(k)$ given by 
\eqref{eq:unified-channel-covariance}. This 
problem is difficult to solve because it is 
non-convex due to the coupling between the 
variables and the non-linear sensing constraint. 
Further, the optimization variables are mixed 
discrete-continuous.

We apply fractional programming (FP) theory 
\cite{Shen-I.2018,Shen-II.2018} to derive an 
equivalent yet more tractable fractional 
formulation. More precisely, we resort to 
closed-form FP, which relies on a Lagrangian 
dual reformulation of the original problem, 
enabling an algorithm where each iteration 
admits a closed-form update instead of 
requiring the numerical solution of a convex 
subproblem as in the direct FP alternative 
\cite{Shen-I.2018,Shen-II.2018}. The 
Lagrangian dual transform converts the cost 
function of problem \textbf{(P1-mod)} to a 
sum-of-ratio form. 

Specifically, let 
$\bar{\boldsymbol{\tau}}^\text{v} \eqdef 
[\bar{\tau}_{1}^\text{v}, \bar{\tau}_{2}^\text{v}, 
\ldots, \bar{\tau}_{K}^\text{v}]^\trasp \in 
\mathbb{R}^K$ be a vector of auxiliary 
variables, for $\text{v} \in \{\text{p},\text{c}\}$. 
Problem \textbf{(P1-mod)} is equivalent to 
maximizing $\bar{\capa}$ in \eqref{eq:bar-rate} 
with respect to $\bm{W}$, 
$\widetilde{\bm{\Phi}}_\text{R}^\text{p}$, 
$\bm{b}$, $\bm{\Phi}_\text{T}^\text{p}$, 
$\bm{\Phi}_\text{R}^\text{c}$, 
$\bm{\Phi}_\text{T}^\text{c}$, 
$\bar{\boldsymbol{\tau}}^\text{p}$, 
$\bar{\boldsymbol{\tau}}^\text{c}$, subject to 
\eqref{12ini-p}, \eqref{12ini-c}, \eqref{12a}, 
\eqref{12b}, \eqref{12c}, and \eqref{12d}, 
where $\bar{\capa}^\text{v}$ is reported in 
\eqref{eq:rate-over-fp} at the top of the 
next page (see \cite{Shen-II.2018} for a 
constructive proof),
\begin{figure*}[!t]
\normalsize
\begin{equation}
\bar{\capa}^\text{v} \eqdef 
\sum_{k=1}^K \log_2(1+\bar{\tau}_{k}^\text{v}) 
- \sum_{k=1}^K \bar{\tau}_{k}^\text{v} + 
\underbrace{\sum_{k=1}^K 
\frac{(1+\bar{\tau}_{k}^\text{v}) \,
\w_k^\herm \, \H_{\text{1}}^\herm 
\{\bm{\Phi}^\text{v}(k)\}^* 
\Rb^\text{v}(k) \, 
\bm{\Phi}^\text{v}(k) \, 
\H_{\text{1}} \, \w_k}
{\displaystyle\sum_{h=1}^K 
\w_h^\herm \, \H_{\text{1}}^\herm 
\{\bm{\Phi}^\text{v}(k)\}^* 
\Rb^\text{v}(k) \, 
\bm{\Phi}^\text{v}(k) \, 
\H_{\text{1}} \, \w_h + 
\sigma_{n_\text{u}}^2}}_{\text{Sum-of-ratio term}}
\label{eq:rate-over-fp}
\end{equation}
\hrulefill
\end{figure*}
for $\text{v} \in \{\text{p},\text{c}\}$. 
It is important to note that \eqref{eq:rate-over-fp} 
is not a new objective function but rather an 
equivalent reformulation of $\bar{\capa}^\text{v}$ 
in \eqref{eq:bar-rate-v} obtained via the Lagrangian 
dual transform: the two expressions attain the same 
value at the optimal $\bar{\boldsymbol{\tau}}^\text{v}$, 
as confirmed by \eqref{eq_taukopt}, which shows that 
the optimal auxiliary variable coincides with the 
statistical SINR $\bar{\gamma}_k^\text{v}$ in 
\eqref{eq:statistical-sinr}.
The equivalence between the two problems stems from 
the fact that $\bar{\capa}$ is a concave 
differentiable function over 
$\bar{\boldsymbol{\tau}}^\text{p}$ and 
$\bar{\boldsymbol{\tau}}^\text{c}$ when 
$\bm{W}$, $\widetilde{\bm{\Phi}}_\text{R}^\text{p}$, 
$\bm{b}$, $\bm{\Phi}_\text{T}^\text{p}$, 
$\bm{\Phi}_\text{R}^\text{c}$, 
$\bm{\Phi}_\text{T}^\text{c}$ are held fixed. 
In this case, the optimal value of 
$\bar{\tau}_{k}^\text{v}$ is obtained by 
setting $\partial\bar{\capa}/\partial 
\bar{\tau}_{k}^\text{v} = 0$, thus yielding
\begin{equation}
\bar{\tau}_{k,\text{opt}}^\text{v} = 
\bar{\gamma}_{k}^\text{v}
\label{eq_taukopt}
\end{equation}
for $\text{v} \in \{\text{p},\text{c}\}$ and 
$k \in \mathcal{K}$.

On the other hand, when 
$\bar{\boldsymbol{\tau}}^\text{v}$ is kept 
fixed, only the sum-of-ratio term in 
\eqref{eq:rate-over-fp} is involved in the 
optimization of the remaining variables. By 
applying the quadratic transform 
\cite{Shen-I.2018,Shen-II.2018} to the 
fractional term in \eqref{eq:rate-over-fp}, 
we may further recast $\bar{\capa}^\text{v}$ 
as reported in \eqref{eq:capavq} at the top 
of the next page,
\begin{figure*}[!t]
\normalsize
\begin{multline}
\bar{\capa}^\text{v}_\text{q} = 
\sum_{k=1}^K \log_2(1+\bar{\tau}_{k}^\text{v}) 
- \sum_{k=1}^K \bar{\tau}_{k}^\text{v} 
+ \sum_{k=1}^K 2\,\bar{\rho}_{k}^\text{v} 
\sqrt{(1+\bar{\tau}_{k}^\text{v}) \,
\w_k^\herm \, \H_{\text{1}}^\herm 
\{\bm{\Phi}^\text{v}(k)\}^* 
\Rb^\text{v}(k) \, 
\bm{\Phi}^\text{v}(k) \, 
\H_{\text{1}} \, \w_k}
\\ - \sum_{k=1}^K (\bar{\rho}_{k}^\text{v})^2 
\left(\sum_{h=1}^K 
\w_h^\herm \, \H_{\text{1}}^\herm 
\{\bm{\Phi}^\text{v}(k)\}^* 
\Rb^\text{v}(k) \, 
\bm{\Phi}^\text{v}(k) \, 
\H_{\text{1}} \, \w_h + 
\sigma_{n_\text{u}}^2\right)
\label{eq:capavq}
\end{multline}
\hrulefill
\end{figure*}
where $\bar{\boldsymbol{\rho}}^\text{v} \eqdef 
[\bar{\rho}_{1}^\text{v}, \bar{\rho}_{2}^\text{v}, 
\ldots, \bar{\rho}_{K}^\text{v}]^\trasp \in 
\mathbb{R}^K$ is a vector of auxiliary 
variables. Hence, to optimize the variables 
$\bm{W}$, $\widetilde{\bm{\Phi}}_\text{R}^\text{p}$, 
$\bm{b}$, $\bm{\Phi}_\text{T}^\text{p}$, 
$\bm{\Phi}_\text{R}^\text{c}$, 
$\bm{\Phi}_\text{T}^\text{c}$, we can 
equivalently maximize (see 
\cite[Corollary~1]{Shen-II.2018} for the 
formal proof)
\begin{equation}
\bar{\capa}_\text{q} = 
\eta \, \bar{\capa}^\text{p}_\text{q} + 
(1-\eta) \, \bar{\capa}^\text{c}_\text{q}
\label{eq:averate-fpq}
\end{equation}
with respect to $\bm{W}$, 
$\widetilde{\bm{\Phi}}_\text{R}^\text{p}$, 
$\bm{b}$, $\bm{\Phi}_\text{T}^\text{p}$, 
$\bm{\Phi}_\text{R}^\text{c}$, 
$\bm{\Phi}_\text{T}^\text{c}$, 
$\bar{\boldsymbol{\rho}}^\text{p}$, 
$\bar{\boldsymbol{\rho}}^\text{c}$, subject to 
\eqref{12ini-p}, \eqref{12ini-c}, \eqref{12a}, 
\eqref{12b}, \eqref{12c}, and \eqref{12d}, 
where the entries of $\bar{\boldsymbol{\tau}}^\text{v}$ 
are iteratively optimized according to 
\eqref{eq_taukopt}, for 
$\text{v} \in \{\text{p},\text{c}\}$. When all 
the other variables are fixed, the optimal 
value of $\bar{\rho}_k^\text{v}$ is obtained 
by setting $\partial\bar{\capa}^\text{v}_\text{q}/
\partial\bar{\rho}_{k}^\text{v} = 0$, i.e.,
\begin{equation}
\bar{\rho}_{k,\text{opt}}^\text{v} = 
\frac{\sqrt{(1+\bar{\tau}_{k}^\text{v}) \,
\w_k^\herm \, \H_{\text{1}}^\herm 
\{\bm{\Phi}^\text{v}(k)\}^* 
\Rb^\text{v}(k) \, 
\bm{\Phi}^\text{v}(k) \, 
\H_{\text{1}} \, \w_k}}
{\displaystyle\sum_{h=1}^K 
\w_h^\herm \, \H_{\text{1}}^\herm 
\{\bm{\Phi}^\text{v}(k)\}^* 
\Rb^\text{v}(k) \, 
\bm{\Phi}^\text{v}(k) \, 
\H_{\text{1}} \, \w_h + 
\sigma_{n_\text{u}}^2}.
\label{eq_rhokopt}
\end{equation}
In the forthcoming subsections, we develop 
the optimization of variables $\bm{W}$, 
$\widetilde{\bm{\Phi}}_\text{R}^\text{p}$, 
$\bm{b}$, $\bm{\Phi}_\text{T}^\text{p}$, 
$\bm{\Phi}_\text{R}^\text{c}$, 
$\bm{\Phi}_\text{T}^\text{c}$ in the 
preparation and communication phases, by 
fixing $\bar{\boldsymbol{\tau}}^\text{v}$ and 
$\bar{\boldsymbol{\rho}}^\text{v}$, for 
$\text{v} \in \{\text{p},\text{c}\}$.

\subsection{Optimization of the beamforming matrix at the BS}
\label{sec:beamformer-opt}

This subsection focuses on optimizing the BS 
beamformer, while keeping 
$\bm{\Phi}_\text{R}^\text{v}$, 
$\bm{\Phi}_\text{T}^\text{v}$, 
$\bar{\boldsymbol{\tau}}^\text{v}$, 
$\bar{\boldsymbol{\rho}}^\text{v}$ fixed, for 
a given stage $\text{v} \in \{\text{p},\text{c}\}$. 
The problem to be solved reads as
\begin{subequations}
\begin{align}
\textbf{(P2)}: \quad & 
\max_{\mathbf{W}} \bar{\capa}^\text{v}_\text{q} 
\label{P3} \\
\text{s.t.} \quad 
& \eta \, \mathbb{E}[\mathrm{SSNR}(\db_k)] \ge 
\delta_\text{sens}, \; 
k \in \mathcal{K}_\text{R} \label{P3a} \\ 
& \mathcal{P} \le \mathcal{P}_{\text{max}} 
\label{P3b}
\end{align}
\end{subequations}
where the transformed objective 
$\bar{\capa}^\text{v}_\text{q}$ is concave in 
$\mathbf{W}$. According to \eqref{eq:Prad}, 
the convex constraint \eqref{P3b} is
\begin{equation}
\sum_{h=1}^K \mathbf{w}_h^\herm \, 
\mathbf{w}_h \le \mathcal{P}_{\text{max}}.
\label{eq:pmax-constr}
\end{equation}
It is worth noting that the beamforming matrix 
has to be optimized in both stages of the 
protocol: during the preparation phase, where 
it jointly supports communication and DoA 
estimation, and again in the communication 
phase, where the updated angular information 
is exploited to refine the communication 
process.

During the communication phase, constraint 
\eqref{P3a} is removed and, thus, problem 
\textbf{(P2)} reduces to a communication-only 
beamforming design under a sum transmit power 
constraint \cite{Tse-book}. In this stage, 
the resulting problem is convex and satisfies 
Slater's condition \cite{Boyd,Beck}. 
Consequently, strong duality holds and the 
Karush--Kuhn--Tucker (KKT) conditions are 
necessary and sufficient for optimality.

The solution of problem \textbf{(P2)} is more 
challenging during the preparation phase, when 
constraint \eqref{P3a} is active. In this 
stage, relying on \eqref{ave-11}, constraint 
\eqref{P3a} can be rewritten as
\begin{equation}
f_k(\mathbf{W}) \eqdef \sum_{h=1}^K 
\mathbf{w}_h^\herm \, \mathbf{P}_k \, 
\mathbf{w}_h \ge 
\frac{\delta_\text{sens} \, N_\text{s} \, 
\sigma_{\text{s},k,\text{eff}}^2}{\eta \, |\alpha_k|^2}
\label{eq:constr-ss-reform}
\end{equation}
with $\mathbf{P}_k \eqdef \H_{\text{1}}^\herm 
\{\bm{\Phi}_{\text{R}}^{\text{p}}\}^* 
\Rb_{\bm{a}_{\text{S}}\bm{a}_{\text{STAR}},k}^\text{p} 
\bm{\Phi}_{\text{R}}^{\text{p}} \H_{\text{1}} 
\in \mathbb{C}^{M \times M}$, which is active 
only during the preparation phase. The sensing 
constraint in \eqref{eq:constr-ss-reform} is 
non-convex because it involves a super-level 
set of a convex quadratic form. To handle 
this, we adopt a successive convex 
approximation (SCA) strategy 
\cite{Razaviyayn,Beck}.

At iteration $\imath \in \mathbb{N}_0$, let 
$\mathbf{W}^\imath \eqdef [\mathbf{w}_1^{\imath}, 
\mathbf{w}_2^{\imath}, \ldots, 
\mathbf{w}_K^{\imath}]$ be the current 
beamformer matrix. The first-order Taylor 
expansion of $f_k(\mathbf{W})$ around 
$\mathbf{W}^\imath$ gives a global affine 
lower bound
\begin{equation}
f_k(\mathbf{W}) \ge 
\sum_{h=1}^K 2\,\Re\!\left\{
[\mathbf{w}_h^{\imath}]^\herm \, 
\mathbf{P}_k \, \mathbf{w}_h\right\}
- \sum_{h=1}^K 
[\mathbf{w}_h^{\imath}]^\herm \, 
\mathbf{P}_k \, \mathbf{w}_h^{\imath}.
\end{equation}
Therefore, at iteration $\imath+1$, we may 
replace \eqref{eq:constr-ss-reform} with its 
affine inner approximation
\begin{equation}
\sum_{h=1}^K 2\,\Re\!\left\{
[\mathbf{w}_h^{\imath}]^\herm \, 
\mathbf{P}_k \, \mathbf{w}_h\right\}
- \sum_{h=1}^K 
[\mathbf{w}_h^{\imath}]^\herm \, 
\mathbf{P}_k \, \mathbf{w}_h^{\imath} 
\ge \frac{\delta_\text{sens} \, N_\text{s} \, 
\sigma_{\text{s},k,\text{eff}}^2}{\eta \, |\alpha_k|^2}
\label{eq:constr-ss-reform-affine}
\end{equation}
for $k \in \mathcal{K}_\text{R}$. Such a set 
of constraints is convex and still 
inner-approximates the original feasible set, 
ensuring monotonic improvement under standard 
SCA conditions \cite{Razaviyayn}. 
Consequently, at iteration $\imath+1$, the 
maximization of $\bar{\capa}^\text{p}_\text{q}$ 
with respect to $\mathbf{W}$, subject to 
\eqref{eq:pmax-constr} and 
\eqref{eq:constr-ss-reform-affine}, is a 
convex quadratically constrained quadratic 
program (QCQP), which can be handled with 
standard solvers \cite{Boyd}. The 
corresponding solution $\mathbf{W}^{\imath+1}$ 
is then updated and the algorithm is repeated 
until convergence.

\subsection{Optimization of the STAR-RIS 
partition during the preparation phase}
\label{sec:part-opt}

In this subsection, we tackle the optimization 
of the binary selection vector $\mathbf{b}$ 
during the preparation phase, when $\mathbf{W}$, 
$\widetilde{\bm{\Phi}}_\text{R}^\text{p}$, 
$\bm{\Phi}_\text{T}^\text{p}$, 
$\bar{\boldsymbol{\tau}}^\text{p}$, 
$\bar{\boldsymbol{\rho}}^\text{p}$ are kept 
fixed. We recall that, during the communication 
phase, all the elements of the STAR-RIS work in 
ES mode, i.e., $\mathbf{b} = \mathbf{1}_{N}$. 
In the preparation stage, the optimization 
problem assumes the form
\begin{subequations}
\begin{align}
\textbf{(P3)}: \quad & 
\max_{\mathbf{b} \in \{0,1\}^N} 
\bar{\capa}^\text{p}_\text{q} \label{P4} \\
\text{s.t.} \quad & 
\|\bm{b}\|_1 = N_\text{part} \le N.
\label{P4a}
\end{align}
\end{subequations}

Problem \textbf{(P3)} is a purely 
combinatorial problem and, therefore, SCA 
cannot be applied directly. To circumvent 
this, we first relax the binary constraint 
by replacing $\mathbf{b} \in \{0,1\}^N$ with 
its continuous approximation 
$\mathbf{b} \in [0,1]^N$ and then enforce a 
``binary-like'' behavior via a penalty term
\begin{equation}
\chi \eqdef \sum_{n=1}^N (b_n - b_n^2)
\end{equation}
which is minimized (to zero) when 
$b_n \in \{0,1\}$ and is maximized when 
$b_n = 1/2$. Subtracting $\chi$ from the 
objective therefore penalizes fractional 
values and drives the solution toward the 
binary feasible set. Specifically, we 
consider the penalized objective
\begin{equation}
\bar{\capa}^\text{p}_{\text{q},\text{pen}} 
\eqdef \bar{\capa}^\text{p}_\text{q} - 
\kappa \, \chi
\end{equation}
which is maximized with respect to 
$\mathbf{b} \in [0,1]^N$, where the regularization parameter 
$\kappa > 0$ controls
the trade-off between objective maximization 
and binary enforcement.

Such an optimization problem is non-convex, 
since $\chi$ is the difference of a linear 
and a convex quadratic function, and is 
solved by leveraging SCA. At iteration 
$\imath+1$, we linearize the concave part 
$-b_n^2$ of $\chi$ around the current point 
$b_n^\imath$ via first-order Taylor expansion:
\begin{equation}
-b_n^2 \approx -(b_n^\imath)^2 - 
2\,b_n^\imath\,(b_n - b_n^\imath)
\end{equation}
which yields the affine approximation of the 
penalty term
\begin{equation}
\chi \approx \sum_{n=1}^N \left[b_n - 
(b_n^\imath)^2 - 
2\,b_n^\imath\,(b_n - b_n^\imath)\right].
\label{eq:varsigma-approx}
\end{equation}
With this linearization, the subproblem at 
iteration $\imath+1$, which amounts to 
maximizing $\bar{\capa}^\text{p}_{\text{q},
\text{pen}}$ with respect to $\mathbf{b}$ 
using the approximation 
\eqref{eq:varsigma-approx}, becomes convex 
and can be tackled with a standard convex 
solver \cite{Boyd}. The resulting solution 
$\bm{b}^{\imath+1} \eqdef [b_1^{\imath+1}, 
b_2^{\imath+1}, \ldots, 
b_N^{\imath+1}]^\trasp \in [0,1]^N$ is then 
updated and the process is repeated until 
convergence, with monotonic improvement 
guaranteed under SCA conditions 
\cite{Razaviyayn}.

Finally, to enforce constraint \eqref{P4a} 
and binarize the vector obtained after 
convergence, the binary partition vector 
$\mathbf{b} \in \{0,1\}^N$ is recovered by 
selecting the $N_\text{part}$ entries of 
largest magnitude and setting them to one, 
while the remaining entries are set to zero. 
This operation constitutes the Euclidean 
projection onto the feasible binary set with 
fixed cardinality.

\subsection{Optimization of the STAR-RIS 
coefficients}
\label{sec:meta-opt}

In this subsection, we tackle the optimization 
of the STAR-RIS parameters while maintaining 
$\mathbf{W}$, $\bm{b}$, 
$\bar{\boldsymbol{\tau}}^\text{v}$, and 
$\bar{\boldsymbol{\rho}}^\text{v}$ fixed, for 
a given stage 
$\text{v} \in \{\text{p},\text{c}\}$. We have 
to solve the following optimization problem
\begin{subequations}
\begin{align}
\textbf{(P4)}: \quad & 
\max_{\widetilde{\bm{\Phi}}_\text{R}^\text{v}, 
\bm{\Phi}_\text{T}^\text{v}} 
\bar{\capa}^\text{v}_\text{q} \label{P5} \\
\text{s.t.} \quad
& \theta_{\text{R},n}^\text{v}, 
\theta_{\text{T},n}^\text{v} \in [0,2\pi), \;
\cos(\theta_{\text{R},n}^\text{v} - 
\theta_{\text{T},n}^\text{v}) = 0 \notag \\
& \beta_{\text{R},n}^\text{v},
\beta_{\text{T},n}^\text{v} \in [0,1], \;
(\beta_{\text{R},n}^\text{v})^2 + 
(\beta_{\text{T},n}^\text{v})^2 = 1, \;
\forall n \in \mathcal{N} \label{12a-P5} \\
& \eta \, \mathbb{E}[\mathrm{SSNR}(\db_k)] \ge 
\delta_\text{sens}, \; 
k \in \mathcal{K}_\text{R} \quad 
(\text{v}=\text{p only}) \label{12b-P5}
\end{align}
\end{subequations}
with $\widetilde{\bm{\Phi}}_\text{R}^\text{c} 
= \bm{\Phi}_\text{R}^\text{c}$ as defined in 
\eqref{eq:TC-RC-matrix}, so that for 
$\text{v}=\text{p}$ the optimization variable 
is $\widetilde{\bm{\Phi}}_\text{R}^\text{p}$ 
(the physical reflection coefficients, 
independently of $\bm{b}$), whereas for 
$\text{v}=\text{c}$ it reduces to 
$\bm{\Phi}_\text{R}^\text{c}$.

The transformed objective function 
$\bar{\capa}^\text{v}_\text{q}$ in 
\eqref{eq:capavq} depends on the STAR-RIS 
coefficients only through quadratic forms of 
the reflection and transmission coefficient 
vectors. Specifically, let
\begin{align}
\boldsymbol{\phi}_{\text{R}}^{\text{p}} 
&\eqdef 
[b_1\widetilde{\beta}_{\text{R},1}^\text{p}
e^{j\theta_{\text{R},1}^\text{p}}, \ldots, 
b_N\widetilde{\beta}_{\text{R},N}^\text{p}
e^{j\theta_{\text{R},N}^\text{p}}]^\trasp \\
\boldsymbol{\phi}_{\text{T}}^{\text{p}} 
&\eqdef 
[\beta_{\text{T},1}^\text{p}
e^{j\theta_{\text{T},1}^\text{p}}, \ldots, 
\beta_{\text{T},N}^\text{p}
e^{j\theta_{\text{T},N}^\text{p}}]^\trasp \\
\boldsymbol{\phi}_{\text{a}}^{\text{c}} 
&\eqdef 
[\beta_{\text{a},1}^\text{c}
e^{j\theta_{\text{a},1}^\text{c}}, \ldots, 
\beta_{\text{a},N}^\text{c}
e^{j\theta_{\text{a},N}^\text{c}}]^\trasp
\end{align}
for $\text{a} \in \{\text{T},\text{R}\}$. By 
using the cyclic property of the trace 
operator \cite{Horn} and element-wise matrix 
multiplication,\footnote{For vectors 
$\mathbf{x}$ and $\mathbf{y}$, and 
corresponding diagonal matrices 
$\mathbf{D}_\mathbf{x}$ and 
$\mathbf{D}_\mathbf{y}$ with these vectors 
as their main diagonals, it holds (see, e.g., 
\cite{Horn}) that $\mathbf{x}^\herm 
(\mathbf{A}\odot\mathbf{B})\mathbf{y} = 
\mathrm{tr}(\mathbf{D}_\mathbf{x}^* \mathbf{A} 
\mathbf{D}_\mathbf{y} \mathbf{B}^\trasp)$.} 
the objective function in \eqref{eq:capavq} 
can be written as reported in 
\eqref{eq:capavq-2} at the top of the next 
page,
\begin{figure*}[!t]
\normalsize
\begin{multline}
\bar{\capa}^\text{v}_\text{q} = 
\sum_{k=1}^K \log_2(1+\bar{\tau}_k^\text{v}) 
- \sum_{k=1}^K \bar{\tau}_k^\text{v}
+ \sum_{k=1}^{K_\text{T}} 
2\bar{\rho}_k^\text{v}
\sqrt{(1+\bar{\tau}_k^\text{v})\,
\{\boldsymbol{\phi}_\text{T}^\text{v}\}^\herm 
\mathbf{E}_{k,k}^\text{v} 
\boldsymbol{\phi}_\text{T}^\text{v}}
+ \sum_{k=K_\text{T}+1}^K 
2\bar{\rho}_k^\text{v}
\sqrt{(1+\bar{\tau}_k^\text{v})\,
\{\boldsymbol{\phi}_\text{R}^\text{v}\}^\herm 
\mathbf{E}_{k,k}^\text{v} 
\boldsymbol{\phi}_\text{R}^\text{v}}
\\ - \sum_{k=1}^{K_\text{T}} 
(\bar{\rho}_k^\text{v})^2\!\left(
\sum_{h=1}^K 
\{\boldsymbol{\phi}_\text{T}^\text{v}\}^\herm 
\mathbf{E}_{k,h}^\text{v} 
\boldsymbol{\phi}_\text{T}^\text{v} 
+ \sigma_{n_\text{u}}^2\right)
- \sum_{k=K_\text{T}+1}^K 
(\bar{\rho}_k^\text{v})^2\!\left(
\sum_{h=1}^K 
\{\boldsymbol{\phi}_\text{R}^\text{v}\}^\herm 
\mathbf{E}_{k,h}^\text{v} 
\boldsymbol{\phi}_\text{R}^\text{v} 
+ \sigma_{n_\text{u}}^2\right)
\label{eq:capavq-2}
\end{multline}
\hrulefill
\end{figure*}
where $\mathbf{E}_{k,h}^\text{v} \eqdef 
\Rb^\text{v}(k) \odot (\H_1 \w_h \w_h^\herm 
\H_1^\herm)^\trasp \in \mathbb{C}^{N\times N}$. 
Similarly, the ASSNR in \eqref{ave-11} can 
be rewritten as
\begin{equation}
\mathbb{E}[\mathrm{SSNR}(\db_k)] = 
\frac{|\alpha_k|^2}
{N_\text{s}\,\sigma_{\text{s},k,\text{eff}}^2}
\{\boldsymbol{\phi}_\text{R}^\text{p}\}^\herm 
\mathbf{D}_k^\text{p} 
\boldsymbol{\phi}_\text{R}^\text{p}, \quad 
k \in \mathcal{K}_\text{R}
\label{ave-11-Hadamard}
\end{equation}
with $\mathbf{D}_k^\text{p} \eqdef 
\Rb_{\bm{a}_\text{S}\bm{a}_\text{STAR},k}^\text{p} 
\odot (\H_1\bm{W}\bm{W}^\herm\H_1^\herm)^\trasp 
\in \mathbb{C}^{N\times N}$. Since 
$\mathbf{D}_k^\text{p}$ is a positive 
semidefinite (PSD) matrix, the quadratic form 
in \eqref{ave-11-Hadamard} is convex in 
$\boldsymbol{\phi}_\text{R}^\text{p}$. 
However, its upper level set is not 
necessarily convex. As a result, problem 
\textbf{(P4)} is non-convex and requires 
additional techniques to be 
solved.

To convexify the problem, we follow a 
semidefinite relaxation (SDR) approach 
\cite{Boyd,Beck} by introducing the rank-one 
PSD matrix $\mathbf{V}_\text{a}^\text{v} 
\eqdef \boldsymbol{\phi}_\text{a}^\text{v} 
\{\boldsymbol{\phi}_\text{a}^\text{v}\}^\herm 
\succeq 0$, for $\text{a} \in \{\text{T},\text{R}\}$ 
and $\text{v} \in \{\text{p},\text{c}\}$. As 
a consequence, using again the cyclic property 
of the trace operator, the quadratic terms in 
\eqref{eq:capavq-2} and 
\eqref{ave-11-Hadamard} become 
$\{\boldsymbol{\phi}_\text{a}^\text{v}\}^\herm 
\mathbf{E}_{k,h}^\text{v} 
\boldsymbol{\phi}_\text{a}^\text{v} = 
\mathrm{tr}(\mathbf{E}_{k,h}^\text{v} 
\mathbf{V}_\text{a}^\text{v})$ and 
$\{\boldsymbol{\phi}_\text{R}^\text{p}\}^\herm 
\mathbf{D}_k^\text{p} 
\boldsymbol{\phi}_\text{R}^\text{p} = 
\mathrm{tr}(\mathbf{D}_k^\text{p} 
\mathbf{V}_\text{R}^\text{p})$, respectively, 
which are linear in 
$\mathbf{V}_\text{a}^\text{v}$.

The STAR-RIS physical constraints can be 
translated into affine constraints in the 
lifted variables. Indeed, the diagonal 
entries of $\mathbf{V}_\text{a}^\text{v}$ 
represent squared amplitudes 
$\{\mathbf{V}_\text{a}^\text{v}\}_{n,n} = 
(\beta_{\text{a},n}^\text{v})^2$ for each 
$n \in \mathcal{N}$. Since the equality 
constraint $(\beta_{\text{R},n}^\text{v})^2 
+ (\beta_{\text{T},n}^\text{v})^2 = 1$ is 
non-convex when combined with the rank-one 
requirement, we adopt the convex relaxation 
$(\beta_{\text{R},n}^\text{v})^2 + 
(\beta_{\text{T},n}^\text{v})^2 \le 1$ for 
each $n \in \mathcal{N}$ and 
$\text{v} \in \{\text{p},\text{c}\}$, which 
can be compactly written as
\begin{equation}
\mathrm{diag}(\mathbf{V}_\text{T}^\text{v}) + 
\mathrm{diag}(\mathbf{V}_\text{R}^\text{v}) 
\le \mathbf{1}_N
\end{equation}
where the inequality is understood 
element-wise.

Since the term 
$\mathrm{tr}(\mathbf{E}_{k,k}^\text{v} 
\mathbf{V}_\text{a}^\text{v})$ in 
\eqref{eq:capavq-2} appears under a square 
root, for $\text{a} \in \{\text{T},\text{R}\}$ 
and $\text{v} \in \{\text{p},\text{c}\}$, we 
introduce auxiliary variables 
$z_{\text{a},k}^\text{v} \ge 0$ with 
$(z_{\text{a},k}^\text{v})^2 \le 
\mathrm{tr}(\mathbf{E}_{k,k}^\text{v} 
\mathbf{V}_\text{a}^\text{v})$, which is a 
rotated second-order cone (SOC) constraint 
\cite{Boyd}. Consequently, the objective 
\eqref{eq:capavq-2} is converted into a 
linear function of $z_{\text{a},k}^\text{v}$ 
and trace-linear functions of the lifted 
matrix $\mathbf{V}_\text{a}^\text{v}$, i.e.,
\begin{multline}
\bar{\capa}^\text{v}_\text{q} = 
\sum_{k=1}^{K_\text{T}} 
2\bar{\rho}_k^\text{v}
\sqrt{1+\bar{\tau}_k^\text{v}}\,
z_{\text{T},k}^\text{v}
+ \sum_{k=K_\text{T}+1}^K 
2\bar{\rho}_k^\text{v}
\sqrt{1+\bar{\tau}_k^\text{v}}\,
z_{\text{R},k}^\text{v}
\\ - \sum_{k=1}^{K_\text{T}} 
\mathrm{tr}(\mathbf{C}_k^\text{v} 
\mathbf{V}_\text{T}^\text{v})
- \sum_{k=K_\text{T}+1}^K 
\mathrm{tr}(\mathbf{C}_k^\text{v} 
\mathbf{V}_\text{R}^\text{v})
+ \mathrm{const}(\bar{\boldsymbol{\tau}}^\text{v},
\bar{\boldsymbol{\rho}}^\text{v})
\end{multline}
where
\begin{equation}
\mathbf{C}_k^\text{v} \eqdef 
(\bar{\rho}_k^\text{v})^2 
\sum_{h=1}^K \mathbf{E}_{k,h}^\text{v} 
\in \mathbb{C}^{N\times N}
\end{equation}
and $\mathrm{const}(\bar{\boldsymbol{\tau}}^\text{v},
\bar{\boldsymbol{\rho}}^\text{v})$ denotes a 
constant term when 
$\bar{\boldsymbol{\tau}}^\text{v}$ and 
$\bar{\boldsymbol{\rho}}^\text{v}$ are fixed. 
By virtue of the Schur complement \cite{Horn}, 
the SOC constraint is guaranteed by the 
linear matrix inequality (LMI)
\begin{equation}
\begin{bmatrix}
\mathrm{tr}(\mathbf{E}_{k,k}^\text{v} 
\mathbf{V}_\text{a}^\text{v}) & 
z_{\text{a},k}^\text{v} \\[2pt]
z_{\text{a},k}^\text{v} & 1
\end{bmatrix} \succeq 0, \quad 
z_{\text{a},k}^\text{v} \ge 0.
\label{eq:SOC}
\end{equation}
By relaxing the rank-one constraints on 
$\mathbf{V}_\text{a}^\text{v}$ while 
preserving the PSD condition 
$\mathbf{V}_\text{a}^\text{v} \succeq 0$, 
we obtain the SDR problem
\begin{subequations}
\begin{align}
\textbf{(P4)-sdr}: \quad & 
\max_{\substack{
\{z_{\text{T},k}^\text{v}\}_{k=1}^K,
\{z_{\text{R},k}^\text{v}\}_{k=1}^K \\ 
\mathbf{V}_\text{T}^\text{v},
\mathbf{V}_\text{R}^\text{v}}}
\bar{\capa}^\text{v}_\text{q} 
\label{P5-sdr} \\
\text{s.t.} \quad
& \mathbf{V}_\text{T}^\text{v} \succeq 0, \; 
\mathbf{V}_\text{R}^\text{v} \succeq 0 \\
& \mathrm{diag}(\mathbf{V}_\text{T}^\text{v}) 
+ \mathrm{diag}(\mathbf{V}_\text{R}^\text{v}) 
\le \mathbf{1}_N \notag \\
& \begin{bmatrix}
\mathrm{tr}(\mathbf{E}_{k,k}^\text{v} 
\mathbf{V}_\text{a}^\text{v}) & 
z_{\text{a},k}^\text{v} \\[2pt]
z_{\text{a},k}^\text{v} & 1
\end{bmatrix} \succeq 0, \; 
z_{\text{a},k}^\text{v} \ge 0 \\
& \mathrm{tr}(\mathbf{D}_k^\text{p} 
\mathbf{V}_\text{R}^\text{p}) \ge 
\frac{\delta_\text{sens} N_\text{s} 
\sigma_{\text{s},k,\text{eff}}^2}
{\eta \, |\alpha_k|^2}, \; 
k \in \mathcal{K}_\text{R}
\label{12b-P5-sdr}
\end{align}
\end{subequations}
This problem is a standard semidefinite 
program (SDP) with LMI and affine constraints 
and can be solved efficiently via 
interior-point methods with polynomial 
complexity \cite{Boyd}. Let 
$\{z_{\text{a},k}^{\text{v}\star}\}_{k=1}^K$ 
and $\mathbf{V}_\text{a}^{\text{v}\star}$ 
denote a solution of \textbf{(P4)-sdr}, for 
$\text{a} \in \{\text{T},\text{R}\}$ and 
$\text{v} \in \{\text{p},\text{c}\}$. Two 
cases may occur. If 
$\mathrm{rank}(\mathbf{V}_\text{a}^{\text{v}\star}) 
= 1$, then $\mathbf{V}_\text{a}^{\text{v}\star} 
= \boldsymbol{\phi}_\text{a}^{\text{v}\star} 
\{\boldsymbol{\phi}_\text{a}^{\text{v}\star}\}^\herm$, 
and the optimal STAR-RIS coefficients are 
obtained directly from the principal 
eigenvector of 
$\mathbf{V}_\text{a}^{\text{v}\star}$. If 
$\mathrm{rank}(\mathbf{V}_\text{a}^{\text{v}\star}) 
> 1$, the relaxation is not tight. A feasible 
rank-one approximation can be constructed via 
principal eigenvector extraction followed by 
normalization, or Gaussian randomization with 
projection onto the feasible STAR-RIS set.

After recovery of the STAR-RIS parameters, 
the phase-coupling constraint 
$\cos(\theta_{\text{R},n}^\text{v} - 
\theta_{\text{T},n}^\text{v}) = 0$ can be 
enforced by projecting the obtained phases 
onto orthogonal pairs, as explained in the 
forthcoming subsection.

\subsection{Phase-coupling restoration via 
euclidean projection}
\label{sec:phase-coup}

Since the optimal solution of \textbf{(P4)-sdr} 
does not necessarily preserve the phase-coupling 
structure, we enforce it by projecting the 
obtained solution onto the feasible STAR-RIS set. 
This is achieved by minimizing the Euclidean 
distance between the normalized reference vector 
and the feasible coefficients:
\begin{subequations}
\begin{align}
\textbf{(P5)}: \quad &
\min_{\bm{\Phi}_\text{R}^\text{v}, 
\bm{\Phi}_\text{T}^\text{v}} \;
\|\bm{\Phi}_\text{R}^{\text{v}\star} - 
\bm{\Phi}_\text{R}^\text{v}\|_2^2 + 
\|\bm{\Phi}_\text{T}^{\text{v}\star} - 
\bm{\Phi}_\text{T}^\text{v}\|_2^2 \\
\text{s.t.} \quad
& \theta_{\text{R},n}^\text{v}, 
\theta_{\text{T},n}^\text{v} \in [0,2\pi), \;
\cos(\theta_{\text{R},n}^\text{v} - 
\theta_{\text{T},n}^\text{v}) = 0 
\label{P7a} \\
& \beta_{\text{R},n}^\text{v}, 
\beta_{\text{T},n}^\text{v} \in [0,1], \;
(\beta_{\text{R},n}^\text{v})^2 + 
(\beta_{\text{T},n}^\text{v})^2 = 1 
\label{P7b} \\
& \forall n \in \mathcal{N} \notag
\end{align}
\end{subequations}
where the diagonal matrix 
$\bm{\Phi}_\text{a}^{\text{v}\star} = 
\mathrm{diag}(\boldsymbol{\phi}_\text{a}^{\text{v}\star})$ 
is obtained from the solution of 
\textbf{(P4)-sdr}. Constraint \eqref{P7a} 
enforces a $\pi/2$ phase difference between 
transmission and reflection coefficients and 
is non-convex due to its discrete nature. 
However, since \eqref{P7a} and \eqref{P7b} 
independently govern phase and amplitude, 
problem \textbf{(P5)} can be solved via 
alternating optimization.

Let the amplitudes $\beta_{\text{R},n}^\text{v}$ 
and $\beta_{\text{T},n}^\text{v}$ be fixed 
for all $n \in \mathcal{N}$. Problem 
\textbf{(P5)} reduces to a projection onto 
the phase-coupled feasible set. The following 
Lagrangian formulation addresses this 
phase-only subproblem. Defining the 
unit-modulus complex variable 
$u_{\text{a},n}^\text{v} \eqdef 
e^{j\theta_{\text{a},n}^\text{v}}$ and the 
unit-modulus complex parameter 
$u_{\text{a},n}^{\text{v}\star} \eqdef 
e^{j\theta_{\text{a},n}^{\text{v}\star}}$ 
(given by the solution of \textbf{(P4)-sdr}), 
for $\text{a} \in \{\text{T},\text{R}\}$ and 
$\text{v} \in \{\text{p},\text{c}\}$, 
constraint \eqref{P7a} is equivalent to the 
orthogonality condition 
$\Re\{u_{\text{R},n}^\text{v} 
(u_{\text{T},n}^\text{v})^*\} = 0$. 
Introducing a real Lagrange multiplier 
$\lambda$, the Lagrangian of the phase-only 
subproblem can be written for all 
$n \in \mathcal{N}$ as
\begin{multline}
\mathcal{L}(u_{\text{R},n}^\text{v}, 
u_{\text{T},n}^\text{v}) \eqdef
|u_{\text{R},n}^{\text{v}\star} - 
u_{\text{R},n}^\text{v}|^2 +
|u_{\text{T},n}^{\text{v}\star} - 
u_{\text{T},n}^\text{v}|^2 
\\ + \lambda\,\Re\{u_{\text{R},n}^\text{v} 
(u_{\text{T},n}^\text{v})^*\}.
\end{multline}
Using Wirtinger calculus \cite{Brandwood}, 
the stationarity conditions are
\begin{align}
\frac{\partial\mathcal{L}}
{\partial(u_{\text{R},n}^\text{v})^*} &= 
(u_{\text{R},n}^{\text{v}\star} - 
u_{\text{R},n}^\text{v}) + 
\frac{\lambda}{2}\,u_{\text{T},n}^\text{v} 
= 0 \label{eq:L1} \\
\frac{\partial\mathcal{L}}
{\partial(u_{\text{T},n}^\text{v})^*} &= 
(u_{\text{T},n}^{\text{v}\star} - 
u_{\text{T},n}^\text{v}) + 
\frac{\lambda}{2}\,u_{\text{R},n}^\text{v} 
= 0. \label{eq:L2}
\end{align}
Let $\mu = -\lambda/2 \in \mathbb{R}$. For 
$\mu^2 \neq 1$, the solution of the system 
\eqref{eq:L1}--\eqref{eq:L2} is given by
\begin{equation}
u_{\text{R},n}^\text{v} = 
\frac{u_{\text{R},n}^{\text{v}\star} - 
\mu\,u_{\text{T},n}^{\text{v}\star}}{1-\mu^2}
\quad \text{and} \quad
u_{\text{T},n}^\text{v} = 
\frac{u_{\text{T},n}^{\text{v}\star} - 
\mu\,u_{\text{R},n}^{\text{v}\star}}{1-\mu^2}.
\label{eq:sol-u}
\end{equation}
Substituting \eqref{eq:sol-u} into the 
orthogonality constraint yields
\begin{equation}
\Re\!\left\{
\frac{(u_{\text{R},n}^{\text{v}\star} - 
\mu\,u_{\text{T},n}^{\text{v}\star})
(u_{\text{T},n}^{\text{v}\star} - 
\mu\,u_{\text{R},n}^{\text{v}\star})^*}
{(1-\mu^2)^2}
\right\} = 0
\end{equation}
which leads to the second-order equation 
$c_1\mu^2 - c_2\mu + c_1 = 0$, with 
$c_1 \eqdef \Re\{u_{\text{R},n}^{\text{v}\star}
(u_{\text{T},n}^{\text{v}\star})^*\}$ and 
$c_2 \eqdef |u_{\text{R},n}^{\text{v}\star}|^2 
+ |u_{\text{T},n}^{\text{v}\star}|^2 = 2$, 
whose solutions are expressed in closed form as
\begin{equation}
\mu = \frac{c_2 \pm \sqrt{c_2^2 - 4c_1^2}}
{2c_1}.
\label{eq:root}
\end{equation}
For each $n \in \mathcal{N}$, the optimal 
phases are then obtained as
\begin{align}
\widehat{\theta}_{\text{R},n}^\text{v} &= 
\arg(u_{\text{R},n}^{\text{v}\star} - 
\mu\,u_{\text{T},n}^{\text{v}\star}) \\
\widehat{\theta}_{\text{T},n}^\text{v} &= 
\arg(u_{\text{T},n}^{\text{v}\star} - 
\mu\,u_{\text{R},n}^{\text{v}\star})
\end{align}
where the root in \eqref{eq:root} minimizing 
the objective is selected. If $c_1 = 0$, then 
$\mu = 0$ and the phases already satisfy the 
orthogonality constraint.

Let us now consider the optimal amplitudes 
for given phases $\theta_{\text{R},n}^\text{v}$ 
and $\theta_{\text{T},n}^\text{v}$, 
$\forall n \in \mathcal{N}$. Problem 
\textbf{(P5)} reduces to
\begin{align}
\max_{\beta_{\text{R},n}^\text{v},
\beta_{\text{T},n}^\text{v} \in [0,1]} \; &
\chi_{\text{R},n}^\text{v}\,
\beta_{\text{R},n}^\text{v} +
\chi_{\text{T},n}^\text{v}\,
\beta_{\text{T},n}^\text{v} \\
\text{s.t.} \quad &
(\beta_{\text{R},n}^\text{v})^2 + 
(\beta_{\text{T},n}^\text{v})^2 = 1 \notag
\end{align}
where $\chi_{\text{a},n}^\text{v} \eqdef 
\beta_{\text{a},n}^{\text{v}\star}
\cos(\theta_{\text{a},n}^{\text{v}\star} - 
\theta_{\text{a},n}^\text{v})$, with 
$\beta_{\text{a},n}^{\text{v}\star}
e^{j\theta_{\text{a},n}^{\text{v}\star}}$ 
obtained by solving \textbf{(P4)-sdr}, for 
$\text{a} \in \{\text{T},\text{R}\}$ and 
$\text{v} \in \{\text{p},\text{c}\}$. This 
amplitude-only subproblem is the maximization 
of a linear function over the unit circle, 
which admits the following closed-form 
solutions:
\begin{itemize}
\item If $\chi_{\text{R},n}^\text{v} \ge 0$ 
and $\chi_{\text{T},n}^\text{v} \ge 0$, then
\[
\widehat{\beta}_{\text{a},n}^\text{v} = 
\frac{\chi_{\text{a},n}^\text{v}}
{\sqrt{(\chi_{\text{R},n}^\text{v})^2 + 
(\chi_{\text{T},n}^\text{v})^2}}, \quad 
\text{a} \in \{\text{T},\text{R}\}.
\]

\item If $\chi_{\text{R},n}^\text{v} \ge 0$ 
and $\chi_{\text{T},n}^\text{v} < 0$, then
\[
\widehat{\beta}_{\text{R},n}^\text{v} = 1 
\quad \text{and} \quad 
\widehat{\beta}_{\text{T},n}^\text{v} = 0.
\]

\item If $\chi_{\text{R},n}^\text{v} < 0$ 
and $\chi_{\text{T},n}^\text{v} \ge 0$, then
\[
\widehat{\beta}_{\text{R},n}^\text{v} = 0 
\quad \text{and} \quad 
\widehat{\beta}_{\text{T},n}^\text{v} = 1.
\]

\item If $\chi_{\text{R},n}^\text{v} < 0$ 
and $\chi_{\text{T},n}^\text{v} < 0$, then
\[
(\widehat{\beta}_{\text{R},n}^\text{v},
\widehat{\beta}_{\text{T},n}^\text{v}) =
\begin{cases}
(0,1), & \chi_{\text{T},n}^\text{v} \ge 
\chi_{\text{R},n}^\text{v} \\
(1,0), & \chi_{\text{R},n}^\text{v} > 
\chi_{\text{T},n}^\text{v}
\end{cases}.
\]
\end{itemize}
We note that when both 
$\chi_{\text{R},n}^\text{v}$ and 
$\chi_{\text{T},n}^\text{v}$ are positive, 
the optimum lies in the interior of the first 
quadrant. Otherwise, the solution occurs at 
the boundary of the feasible set.

\subsection{Overall alternating optimization procedure}

For fixed beamforming matrices and STAR-RIS 
coefficients, the time-allocation factor $\eta$ 
can be updated in closed form. Since the 
recovered feasible solution after relaxation 
and projection may provide an SSNR higher than 
the required threshold, $\eta$ is updated based 
on the actually achieved preparation-stage SSNR. 
Since sensing is only performed during the 
preparation stage, the accumulated sensing 
requirement imposes
\begin{equation}
\eta \, \overline{\text{SSNR}}_k^\text{p} \ge 
\delta_\text{sens}, \quad 
k \in \mathcal{K}_\text{R},
\label{eq:eta-ssnr-constraint}
\end{equation}
where $\overline{\text{SSNR}}_k^\text{p}$ 
denotes the average sensing SNR of the $k$-th 
outdoor user in the preparation stage, 
evaluated at the current beamforming and 
STAR-RIS coefficients via \eqref{ave-11}. 
Hence, the feasible lower bound of $\eta$ is
\begin{equation}
\eta_\text{lb} = \max\!\left\{\eta_{\min}, \,
\max_{k \in \mathcal{K}_\text{R}} 
\frac{\delta_\text{sens}}
{\overline{\text{SSNR}}_k^\text{p}}
\right\}.
\label{eq:eta-lower-bound}
\end{equation}
Let $\bar{\capa}^\text{p} \eqdef 
\mathbb{E}[\capa^\text{p}]$ and 
$\bar{\capa}^\text{c} \eqdef 
\mathbb{E}[\capa^\text{c}]$. Since 
$\mathbb{E}[\capa]$ is affine with respect 
to $\eta$, the updated value is given by
\begin{equation}
\eta^\star = \begin{cases}
\eta_\text{lb}, & 
\bar{\capa}^\text{p} < \bar{\capa}^\text{c}, \\
\eta_\text{max}, & 
\bar{\capa}^\text{p} > \bar{\capa}^\text{c}, \\
[\eta]_{\eta_\text{lb}}^{\eta_\text{max}}, & 
\bar{\capa}^\text{p} = \bar{\capa}^\text{c},
\end{cases}
\label{eq:eta-update}
\end{equation}
where $[\cdot]_{\eta_\text{lb}}^{\eta_\text{max}}$ 
denotes the projection onto 
$[\eta_\text{lb}, \eta_\text{max}]$.

The optimization problems addressed in 
Subsections~III-\ref{sec:beamformer-opt}, 
III-\ref{sec:part-opt}, III-\ref{sec:meta-opt}, and 
III-\ref{sec:phase-coup} provide update rules 
and tractable subroutines for each variable 
block. In this subsection we describe how 
these blocks are coordinated to solve problem 
\textbf{(P1-mod)}. We adopt a block 
coordinate descent (BCD) framework 
\cite{Bertsekas}, where at each iteration 
one block of variables is updated while 
keeping the others fixed. More specifically, 
we resort to a penalty parameter updating 
strategy, widely used in augmented-Lagrangian 
and constrained non-convex optimization 
\cite{Bertsekas-AL,Razaviyayn}, which enables 
the BCD algorithm to first locate a reliable 
operating point in a relaxed feasible region 
and subsequently enforce the physical 
STAR-RIS constraints. This continuation 
mechanism mitigates poor stationary points 
commonly encountered in tightly coupled 
non-convex beamforming problems.

During the preparation stage, the algorithm 
updates the variables sequentially as follows:

\begin{enumerate}
\item Update $\bar{\boldsymbol{\tau}}^\text{p}$ 
and $\bar{\boldsymbol{\rho}}^\text{p}$ using 
\eqref{eq_taukopt} and \eqref{eq_rhokopt}.

\item Update the BS beamforming matrix 
$\mathbf{W}$ by solving problem \textbf{(P2)} 
in \eqref{P3}--\eqref{P3b} via SCA.

\item Update the STAR-RIS partition vector 
$\mathbf{b}$ using the penalized relaxation 
and projection in 
Subsection~\ref{sec:part-opt}.

\item Update the STAR-RIS coefficients 
$(\bm{\Phi}_\text{T}^\text{p},
\bm{\Phi}_\text{R}^\text{p})$ by solving 
the SDR problem \textbf{(P4)-sdr} in 
\eqref{P5-sdr}--\eqref{12b-P5-sdr} and 
applying rank-one recovery.

\item Recover feasible element coefficients 
via the element-wise projection in 
Subsection~\ref{sec:phase-coup}.
\end{enumerate}

During the communication stage, sensing 
constraints are inactive and all elements 
operate in ES mode ($\mathbf{b} = 
\mathbf{1}_N$). The updates reduce to the 
following steps:

\begin{enumerate}
\item Update $\bar{\boldsymbol{\tau}}^\text{c}$ 
and $\bar{\boldsymbol{\rho}}^\text{c}$ using 
\eqref{eq_taukopt} and \eqref{eq_rhokopt}.

\item Update $\mathbf{W}$ by solving the 
convex communication-only beamforming problem 
\cite{Tse-book,Boyd,Beck}.

\item Update $(\bm{\Phi}_\text{T}^\text{c},
\bm{\Phi}_\text{R}^\text{c})$ by solving the 
SDR problem \textbf{(P4)-sdr} in 
\eqref{P5-sdr} and applying rank-one recovery.

\item Recover feasible element coefficients 
via the element-wise projection in 
Subsection~\ref{sec:phase-coup}.
\end{enumerate}

After the updates of both stages, the 
time-allocation factor $\eta$ is updated via 
\eqref{eq:eta-update}. This scalar update 
does not require solving an additional 
optimization subproblem and is performed 
before checking the convergence criterion.

Each block update either (i) maximizes the 
objective with the remaining variables fixed, 
(ii) maximizes a tight surrogate (SCA) that 
lower-bounds the original cost, or (iii) 
updates the scalar time-allocation factor 
$\eta$ via the closed-form solution 
\eqref{eq:eta-update}. Hence, the sequence 
of objective values is monotonically 
non-decreasing and upper bounded, due to the 
finite transmit power and the bounded 
interval $\eta \in [\eta_{\mathrm{min}}, \eta_{\mathrm{max}}]$. 
By standard results for block-coordinate 
methods \cite{Bertsekas,Bertsekas-AL,
Razaviyayn}, the procedure converges to a 
stationary point of the original problem 
under mild regularity conditions.

In practice, the outer loop is terminated 
when the relative improvement of the 
objective between two successive iterations 
falls below a threshold $\epsilon$ (e.g., 
$10^{-3}$) or when a maximum number of 
iterations $I_\text{max}$ is reached. Inner 
subproblems (SDPs/SCA) are solved to 
moderate tolerances to balance accuracy and 
runtime. The computational bottleneck is 
represented by the SDP solvers in problem 
\textbf{(P4)-sdr} 
(Subsection~\ref{sec:meta-opt}), whose 
complexity scales polynomially with the 
STAR-RIS dimension $N$ (roughly cubic for 
interior-point solvers). The SCA and 
projection steps are comparatively 
inexpensive. For large-scale arrays, 
first-order SDP solvers or low-rank 
approximations can be employed; this is 
left for future work.

\subsection{Computational complexity analysis}
\label{sec:complexity}

We briefly analyze the computational complexity 
of the proposed alternating optimization 
algorithm. Let $I_\text{out}$ denote the number 
of outer BCD iterations, and let $I_\text{W}$, 
$I_\text{b}$, and $I_\Phi$ denote the numbers 
of inner iterations required for updating the 
BS beamforming matrix, the STAR-RIS partition 
vector, and the STAR-RIS coefficients, 
respectively.

The closed-form updates of the auxiliary 
variables $\{\bar{\tau}_k^\text{v}\}$ and 
$\{\bar{\rho}_k^\text{v}\}$ mainly involve 
the evaluation of statistical quadratic forms 
and require $\mathcal{O}(K^2NM)$ operations 
per stage. The BS beamforming update solves 
a convex QCQP with $KM$ complex variables, 
whose complexity is approximately 
$\mathcal{O}(I_\text{W}(KM)^3)$. The relaxed 
partition optimization involves $N$ real 
variables and has complexity 
$\mathcal{O}(I_\text{b}N^3)$, followed by a 
projection step with complexity 
$\mathcal{O}(N\log N)$.

The dominant computational burden comes from 
the STAR-RIS coefficient update. Specifically, 
\textbf{(P4)-sdr} optimizes two lifted PSD 
matrices of size $N \times N$, i.e., 
$\mathbf{V}_\text{T}^\text{v}$ and 
$\mathbf{V}_\text{R}^\text{v}$. When an 
interior-point SDP solver is adopted, the 
corresponding complexity scales approximately 
as $\mathcal{O}(I_\Phi N^6)$ \cite{Boyd}. 
The subsequent phase-coupling restoration is 
performed element-wise and requires only 
$\mathcal{O}(N)$ operations. Therefore, the 
overall complexity of the proposed algorithm 
can be summarized as
\begin{multline}
\mathcal{O}\!\Big(I_\text{out}\big[
K^2NM + I_\text{W}(KM)^3 + 
I_\text{b}N^3 + I_\Phi N^6 + 
N\log N\big]\Big).
\label{eq:complexity-total}
\end{multline}
It can be observed from 
\eqref{eq:complexity-total} that the 
SDP-based STAR-RIS coefficient optimization 
is the computational bottleneck, especially 
for large-scale STAR-RISs. This motivates 
the development of low-complexity first-order, 
low-rank, or manifold-based alternatives for 
large-scale real-time implementations, which 
is left for future work.


\begin{table*}[t]
\centering
\caption{Main characteristics of the 
considered benchmark schemes.}
\label{tab:grid_table-1}
\begin{tabular}{lccccc}
\toprule
Scheme & Sensing & Phase& 
Spatial expectation  & 
Manifold & Dual  \\
       & SNR & coupling & 
statistics & 
design & stage \\
\midrule
Proposed & \checkmark  & \checkmark & \checkmark & \checkmark & \checkmark \\
SC-STAR \cite{Liu.2025.TVT} & 
\checkmark & \ding{55} & \ding{55} & \ding{55} & \ding{55} \\
PGAM-STAR \cite{Papazafeiropoulos.2024.TWC} & 
 \checkmark & \ding{55} & \ding{55} & \ding{55} & \ding{55} \\
NoStat-STAR & \checkmark & \checkmark & \ding{55} & \checkmark & \checkmark \\
Fixed-STAR & \checkmark & \checkmark & \checkmark & \ding{55} & \checkmark \\
One-stage & \checkmark & \checkmark & \checkmark & \ding{55} & \ding{55} \\
\bottomrule
\end{tabular}
\end{table*}

\section{Numerical results}
\label{sec:Simulation}

This section evaluates the performance of the 
proposed STAR-RIS-assisted ISAC framework 
through Monte Carlo numerical simulations. The 
objective is to assess the achievable 
communication-sensing trade-off and to quantify 
the benefits of jointly optimizing beamforming 
weights, metasurface coefficients, and element 
partitioning. To this end, the proposed design 
is compared with several benchmark schemes 
representing different modeling assumptions and 
levels of optimization capability. Specifically, 
the following benchmark schemes are considered.

\begin{enumerate}

\item \emph{Statistical-CSI STAR-RIS ISAC 
(SC-STAR)}: This scheme is adapted from the 
statistical-CSI STAR-RIS-enabled ISAC design 
in~\cite{Liu.2025.TVT}. It adopts a 
single-stage all-ES STAR-RIS architecture 
without sensing-assisted angular refinement or 
STAR-RIS partition optimization. The BS 
beamforming matrix and the STAR-RIS T/R 
coefficients are optimized using the 
center-based second-order CSI under the same 
SSNR constraint, and the final sum-rate is 
evaluated over the sampled real channel.

\item \emph{Statistical-CSI STAR-RIS design 
with PGAM (PGAM-STAR)}: This scheme is 
inspired by the statistical-CSI STAR-RIS 
design in~\cite{Papazafeiropoulos.2024.TWC}. 
It adopts a single-stage all-ES architecture, 
where the STAR-RIS T/R amplitudes and phase 
shifts are optimized by the projected gradient 
ascent method (PGAM) using the center-based 
statistical CSI. Unlike the proposed method, 
it does not perform sensing-assisted angular 
refinement or partition optimization; the BS 
beamforming matrix is subsequently refined 
under the same SSNR constraint.

\item \emph{Proposed STAR-RIS without 
statistical information (NoStat-STAR)}: This 
scheme relies on the proposed optimization 
framework but neglects the spatial statistical 
information used in the main design. It is 
introduced to isolate the performance gain 
obtained from exploiting estimation error 
statistics.

\item \emph{Proposed STAR-RIS with fixed 
manifold (Fixed-STAR)}: The metasurface 
partition is predetermined and not optimized. 
Specifically, half of the elements operate in 
ES mode while the remaining half operate in 
TO mode. This scheme evaluates the benefit of 
adaptive element partitioning.

\item \emph{One-stage design (One-stage)}: 
This scheme adopts the same statistical design 
framework and STAR-RIS partition optimization 
as the proposed method, but operates within a 
single transmission stage without 
preparation/communication phase separation. 
It is introduced to isolate the performance 
gain attributable to the two-stage protocol.

\end{enumerate}

The main characteristics of the considered 
benchmark schemes are summarized in 
Table~\ref{tab:grid_table-1}.

\begin{table}[t]
\centering
\caption{Simulation setting.}
\label{tab:grid_table-2}
\begin{tabular}{ll}
\toprule
Parameter & Value \\
\midrule
Number of antennas at the BS & $M = 8$ \\
Number of sensors & $N_\text{s} = 8$ \\
Number of users & $K = 4$ \\
Number of STAR-RIS elements & $N = 20$ \\
Transmit power of BS & 
$\mathcal{P}_\text{max} = 20$~dBm \\
Rician factor & 
$\mu_1 = \mu_{\text{2,in}} = 
\mu_{\text{2,out}} = 2$ \\
Noise power & 
$\sigma_{n_\text{u}}^2 = 
\sigma_{n_\text{s}}^2 = -110$~dBm \\
Effective disturbance variance & 
$\sigma_{\text{s},k,\text{eff}}^2 = 
\sigma_{n_\text{s}}^2$ (no clutter) \\
Path loss at $d_0 = 1$~m & 
$\varsigma_0 = 30$~dB \\
Sensing SNR threshold & 
$\delta_\text{sens} = 10$~dB \\
Number of ES elements & 
$N_\text{part} = 10$ \\
Sensing complex amplitude & 
$\alpha_k = -10$~dB \\
\bottomrule
\end{tabular}
\end{table}

According to Fig.~\ref{fig:3}, the STAR-RIS 
is placed at the origin $(0,0,0)$~m, while 
the BS is located at $(20,30,0)$~m. The 
numbers of users in the transmission and 
reflection regions are assumed to be equal, 
i.e., $K_\text{T} = K_\text{R}$. The random 
position of user $k$ is expressed in 
spherical coordinates as
\begin{equation}
\mathbf{p}_k =
\begin{cases}
(r_k, \phi_k, \varphi_k), & 
k \in \mathcal{K}_\text{T} \\[4pt]
(r_k, \phi_k+\Delta\phi_k, 
\varphi_k+\Delta\varphi_k), & 
k \in \mathcal{K}_\text{R}
\end{cases}
\end{equation}
with $r_k \sim \mathcal{U}[30,50]$~m, 
$\phi_k \sim \mathcal{U}[0^\circ,180^\circ]$, 
and $\varphi_k \sim 
\mathcal{U}[-90^\circ,90^\circ]$, where the 
angular uncertainties are modeled as 
$\Delta\phi_k, \Delta\varphi_k \sim 
\mathcal{N}(0,\mathrm{PCRB})$, with the PCRB 
given in~\cite{Wang.2023.TWC}. The remaining 
system parameters are summarized in 
Table~\ref{tab:grid_table-2}. This baseline 
configuration is kept unchanged across all 
simulation results unless otherwise specified.

\begin{figure}[t]
\centering
\includegraphics[width=1\linewidth]{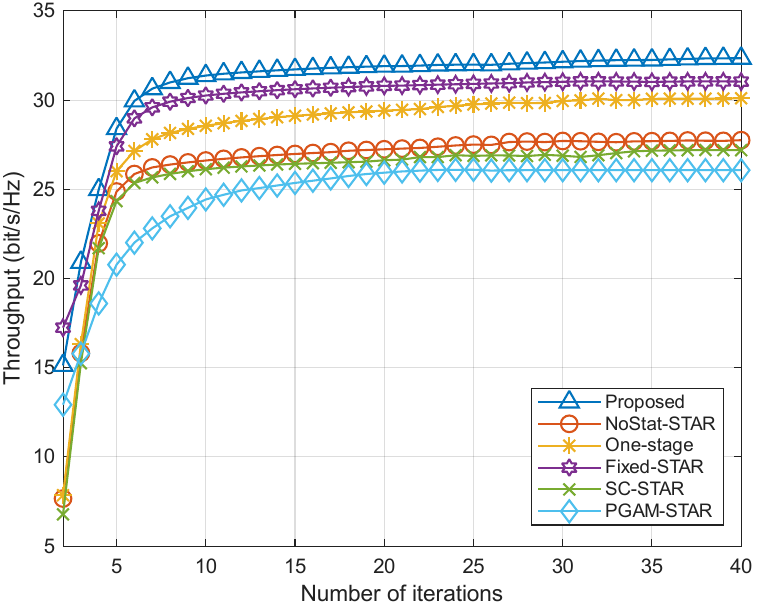}
\caption{Throughput versus number of 
iterations.}
\label{fig:iteration}
\end{figure}

\subsection{Throughput as a function of the 
number of iterations}

The convergence behavior of the proposed 
algorithm is evaluated in terms of the 
achievable throughput, whose evolution over 
the iterations is illustrated in 
Fig.~\ref{fig:iteration}. The curve exhibits 
monotonic convergence within a limited number 
(about $10$) of iterations. In addition, 
results not reported here show that the phase 
difference stabilizes around $\pi/2$ and 
$3\pi/2$ after approximately $15$ iterations. 
As can be observed from 
Fig.~\ref{fig:iteration}, SC-STAR, PGAM-STAR, 
and NoStat-STAR methods provide 
comparable but inferior
performances, since none of them explicitly 
takes into account spatial-domain constraints. 
Among them, NoStat-STAR achieves about $4\%$ 
and $7\%$ performance improvement over 
SC-STAR and PGAM-STAR, respectively. This 
difference is mainly caused by the distinct 
optimization frameworks and constraint designs 
adopted by the considered schemes, which 
result in different operating trade-offs while 
still confirming the feasibility of the 
proposed framework. The performance gap 
becomes more evident when compared with the 
Fixed-STAR scheme, whose fixed architecture 
limits design flexibility and inherently 
constrains the achievable performance. 
Moreover, the restricted effective aperture 
deteriorates estimation accuracy, which may 
further amplify beam misalignment under 
non-ideal conditions. Overall, the results 
show that incorporating spatial-domain 
statistical information yields a $15.6\%$ 
performance gain for the proposed scheme, 
the flexible time-slot allocation provides 
an additional $7.5\%$ gain, and the increased 
design flexibility provides a further $4.3\%$ 
improvement.

\begin{figure}[t]
\centering
\includegraphics[width=1\linewidth]{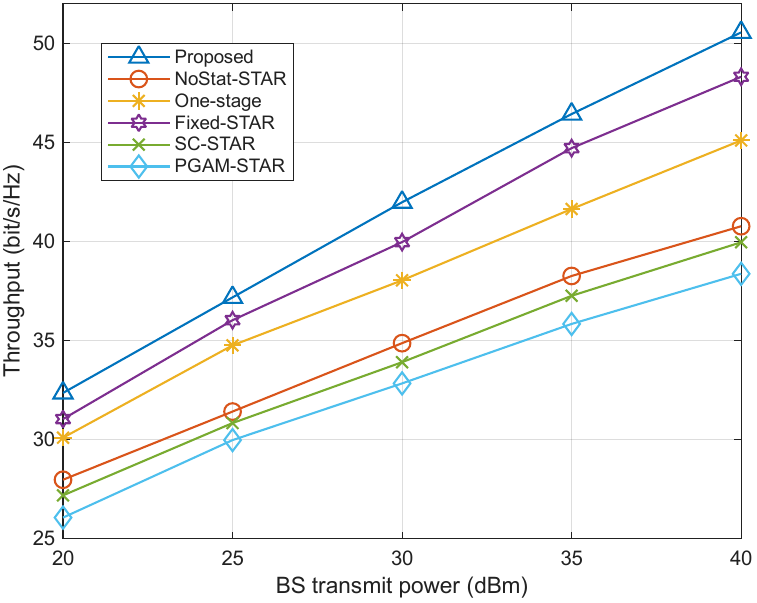}
\caption{Throughput versus BS transmit power.}
\label{fig:pow}
\end{figure}

\subsection{Throughput as a function of the 
BS power budget}

The variation of the achievable throughput 
with respect to the BS transmit power (in dBm) is 
shown in Fig.~\ref{fig:pow}. As expected, 
the achievable rate increases with the 
transmit power, since higher available power 
results in greater signal strength at the 
users. It can be observed that SC-STAR, 
NoStat-STAR, and PGAM-STAR schemes exhibit 
similar performance levels, whereas the 
proposed method consistently achieves about 
$15\%$ higher rate values.

\begin{figure}[t]
\centering
\includegraphics[width=1\linewidth]
{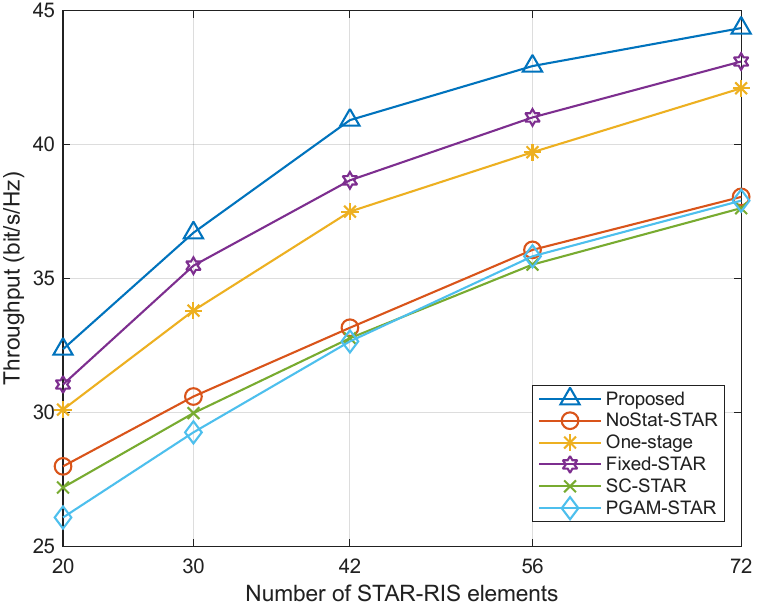}
\caption{Throughput versus  
number of STAR-RIS elements.}
\label{fig:nummeta}
\end{figure}

\subsection{Throughput as a function of the 
number of STAR-RIS elements}

The achievable throughput as a function of 
the number $N$ of STAR-RIS elements is shown 
in Fig.~\ref{fig:nummeta}. For all considered 
schemes, the achievable rate increases as the 
number of metasurface elements grows. The 
proposed method exhibits an improvement of 
approximately $24\%$ for each doubling of 
the element count. This behavior is expected, 
since a larger STAR-RIS provides higher 
beamforming gain. Furthermore, the increased 
aperture enhances DoA estimation accuracy, 
resulting in improved beam alignment. The 
additional design flexibility offered by the 
proposed scheme further contributes to its 
performance advantage.

\begin{figure}[t]
\centering
\includegraphics[width=1\linewidth]{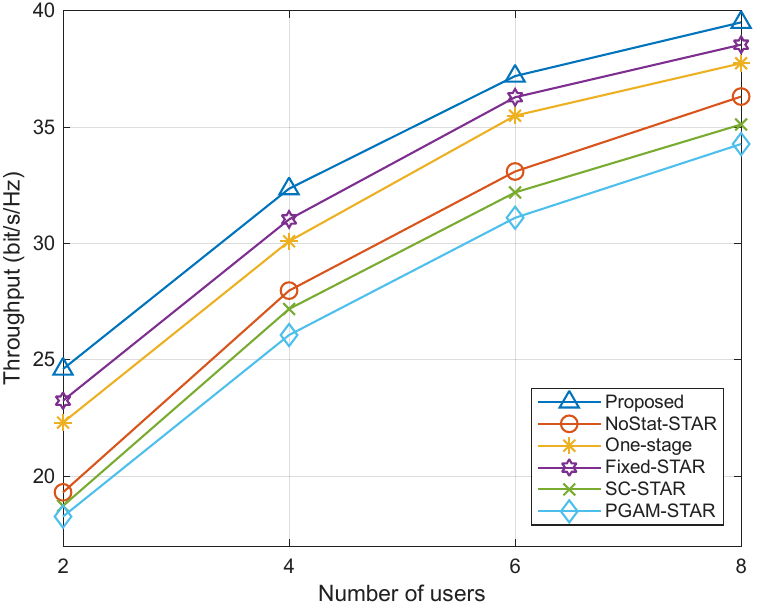}
\caption{Throughput versus number of users.}
\label{fig:numuser}
\end{figure}

\subsection{Throughput as a function of the 
number of users}

The achievable throughput as a function of 
the number of users is shown in 
Fig.~\ref{fig:numuser}. The results are 
obtained under the condition 
$K_\text{T} = K_\text{R}$ and in the 
presence of DoA estimation errors in the 
reflection space. All schemes exhibit 
sublinear growth as the number of users 
increases, due to intensified resource 
competition and the resulting inter-user 
interference. The absence of a minimum 
per-user rate constraint in the formulation 
further accentuates this behavior. The 
proposed method maintains a consistent 
performance advantage over all benchmarks 
across the considered range of user counts, 
confirming the scalability of the proposed 
framework.

\begin{figure}[t]
\centering
\includegraphics[width=1\linewidth]{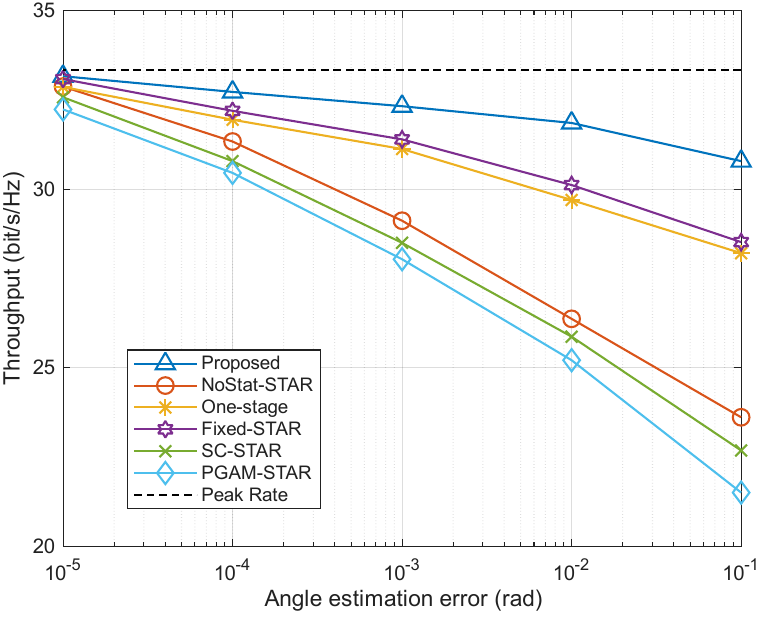}
\caption{Throughput versus DoA estimation 
error.}
\label{fig:numerr}
\end{figure}

\subsection{Throughput as a function of 
DoA estimation error}

The achievable throughput as a function of 
the estimation error is shown in 
Fig.~\ref{fig:numerr}. The peak throughput 
corresponds to the achievable rate under 
perfect angular knowledge. As the estimation 
error decreases, all schemes under comparison 
gradually approach the performance of the 
ideal error-free case. The proposed scheme 
exhibits the slowest performance degradation 
and consistently achieves the highest 
throughput among all practical schemes, 
demonstrating its robustness against DoA 
uncertainty and imperfect NLoS channel 
knowledge.

\subsection{Robustness to equivalent 
sensing-parameter fluctuations}

\begin{table}[t]
\centering
\caption{Impact of equivalent 
sensing-parameter fluctuations on the 
proposed two-stage scheme.}
\label{tab:sensing_mismatch}
\footnotesize

\textbf{Equivalent target-response 
fluctuation}

\vspace{0.5mm}
\begin{tabular*}{\linewidth}
{@{\extracolsep{\fill}}lccccc@{}}
\toprule
$\Delta_{\alpha}$ (dB) & 
$-6$ & $-3$ & $0$ & $+3$ & $+6$ \\
\midrule
$\mathbb{E}[\capa]$ (bit/s/Hz) & 
29.75 & 30.96 & 32.36 & 33.06 & 33.94 \\
Std.\ of $\capa$ & 
1.19 & 1.22 & 1.13 & 1.17 & 1.27 \\
$\mathcal{M}_\text{sens}$ & 
1.0394 & 1.0246 & 1.0000 & 1.0096 & 
1.0151 \\
\bottomrule
\end{tabular*}

\vspace{1.5mm}

\textbf{Residual-scattering disturbance 
fluctuation}

\vspace{0.5mm}
\begin{tabular*}{\linewidth}
{@{\extracolsep{\fill}}lcccc@{}}
\toprule
$\Delta_{\sigma}$ (dB) & 
$0$ & $+3$ & $+6$ & $+9$ \\
\midrule
$\mathbb{E}[\capa]$ (bit/s/Hz) & 
32.36 & 31.01 & 29.06 & 27.92 \\
Std.\ of $\capa$ & 
1.13 & 1.19 & 1.17 & 1.21 \\
$\mathcal{M}_\text{sens}$ & 
1.0000 & 1.0131 & 1.0512 & 1.0326 \\
\bottomrule
\end{tabular*}

\vspace{1mm}
\begin{minipage}{\linewidth}
\scriptsize
$\mathcal{M}_\text{sens} \eqdef 
\min_{k \in \mathcal{K}_\text{R}} 
\,\eta\overline{\mathrm{SSNR}}_k^\text{p} / 
\delta_\text{sens}$. A sensing margin larger 
than one indicates that the sensing 
constraint is satisfied.
\end{minipage}
\end{table}

Table~\ref{tab:sensing_mismatch} evaluates 
the robustness of the proposed two-stage 
scheme to equivalent sensing-parameter 
fluctuations, including target-response 
mismatch (variations in $\alpha_k$) and 
residual-scattering disturbance fluctuations 
(variations in $\sigma_{\text{s},k,\text{eff}}^2$). 
A stronger target response or a weaker 
residual disturbance leads to a larger 
preparation-stage SSNR, thereby reducing 
the sensing burden in the time-allocation 
update. Conversely, increased disturbance 
degrades the SSNR and reduces the achievable 
sum-rate. Nevertheless, the sensing margin 
$\mathcal{M}_\text{sens}$ remains larger 
than one in all considered cases, indicating 
that the sensing constraint \eqref{12b} is 
satisfied under all considered 
sensing-parameter fluctuations. The 
relatively small standard deviations further 
confirm the stability of the observed trends 
over Monte Carlo realizations.

\section{Conclusions}
\label{sec:conclusions}

This paper investigated the robust design of 
STAR-RIS-assisted ISAC systems from a 
protocol-level perspective, addressing two 
fundamental challenges that limit practical 
deployments: DoA estimation uncertainty for 
outdoor users and partial knowledge of NLoS 
channel components. By leveraging the 
full-space transmission and reflection 
capabilities of STAR-RISs, a two-stage ISAC 
protocol was proposed, in which a preparation 
phase jointly supports DoA estimation 
and downlink communication, while a 
subsequent communication phase exploits the 
acquired angular information to enhance 
information transfer to both indoor and 
outdoor users.

Unlike most existing works that assume perfect 
instantaneous NLoS channel knowledge, the 
proposed framework characterizes the 
STAR-RIS-to-outdoor-user links through their 
long-term spatial covariance statistics, 
enabling a robust design that incorporates 
average communication performance into the 
optimization without requiring instantaneous 
NLoS CSI. The DoAs of outdoor users 
were modeled as Gaussian random variables 
whose variances are determined by the 
estimation accuracy, and both sources of 
uncertainty were jointly accounted for in 
the system design.

A performance-balanced optimization problem 
was formulated, jointly optimizing the BS 
beamforming vectors, the STAR-RIS 
transmission and reflection coefficients in 
both stages, and the metasurface partition 
between energy-splitting and transmit-only 
modes, while explicitly enforcing STAR-RIS 
physical feasibility constraints and 
sensing-quality requirements. To tackle the 
resulting non-convex mixed discrete--continuous 
problem, a tailored alternating optimization 
framework was developed, combining fractional 
programming, Lagrangian dual reformulation, 
successive convex approximation, and 
semidefinite relaxation, with proven monotonic 
convergence to a stationary point. The 
computational complexity of the proposed 
algorithm was characterized in closed form, 
identifying the SDP-based STAR-RIS 
coefficient update as the dominant bottleneck.

Numerical results demonstrated that the 
proposed robust design achieves an effective 
sensing-communication trade-off, yielding 
approximately $15\%$ throughput gain over 
the most competitive benchmark -- a scheme 
sharing the same optimization framework but 
neglecting NLoS spatial covariance statistics 
-- under both ideal and imperfect sensing 
conditions. The sensing constraint was shown 
to remain satisfied across all considered 
target-response and residual-clutter 
fluctuations, with sensing margins consistently 
above unity. The robustness of the proposed 
design against DoA estimation errors was 
further confirmed, with the proposed scheme 
exhibiting the slowest performance degradation 
among all benchmarks as the estimation error 
increases.

Future work will investigate the extension 
of the proposed framework to more general 
EM STAR-RIS models incorporating 
mutual coupling and hardware impairments, 
the addition of per-user quality-of-service 
constraints, the development of 
low-complexity first-order or manifold-based 
alternatives to the SDP solver for large-scale 
arrays, and the explicit incorporation of 
Doppler effects for high-mobility scenarios.


\end{document}